# The Dark Energy Spectrometer (DESpec): A Multi-Fiber Spectroscopic Upgrade of the Dark Energy Camera and Survey for the Blanco Telescope

September 11, 2012

Authors:  F. Abdalla (1), J. Annis (2), D. Bacon (3), S. Bridle (1), F. Castander (4), M. Colless (5), D. DePoy (6), H. T. Diehl (2), M. Eriksen (4), B. Flaugher (2), J. Frieman (2, 7), E. Gaztanaga (4), C. Hogan (2, 7), S. Jouvel (4), S. Kent (2, 7), D. Kirk (1), R. Kron (2, 7), S. Kuhlmann (8), O. Lahav (1), J. Lawrence (5), H. Lin (2), J. Marriner (2), J. Marshall (6), J. Mohr (9), R. C. Nichol (3), M. Sako (10), W. Saunders (5), M. Soares-Santos (2), D. Thomas (3), R. Wechsler (11), A. West (2), H. Wu (11) ((1) University College London, UK, (2) Fermilab, (3) Institute of Cosmology and Gravitation, Portsmouth, UK, (4) Institut de Ciències de l'Espai, Barcelona, Spain, (5) Australian Astronomical Observatory, (6) Texas A&M University, (7) University of Chicago, (8) Argonne National Laboratory, (9) Ludwig-Maximilians University, Germany, (10) University of Pennsylvania,  (11) KIPAC, Stanford University)

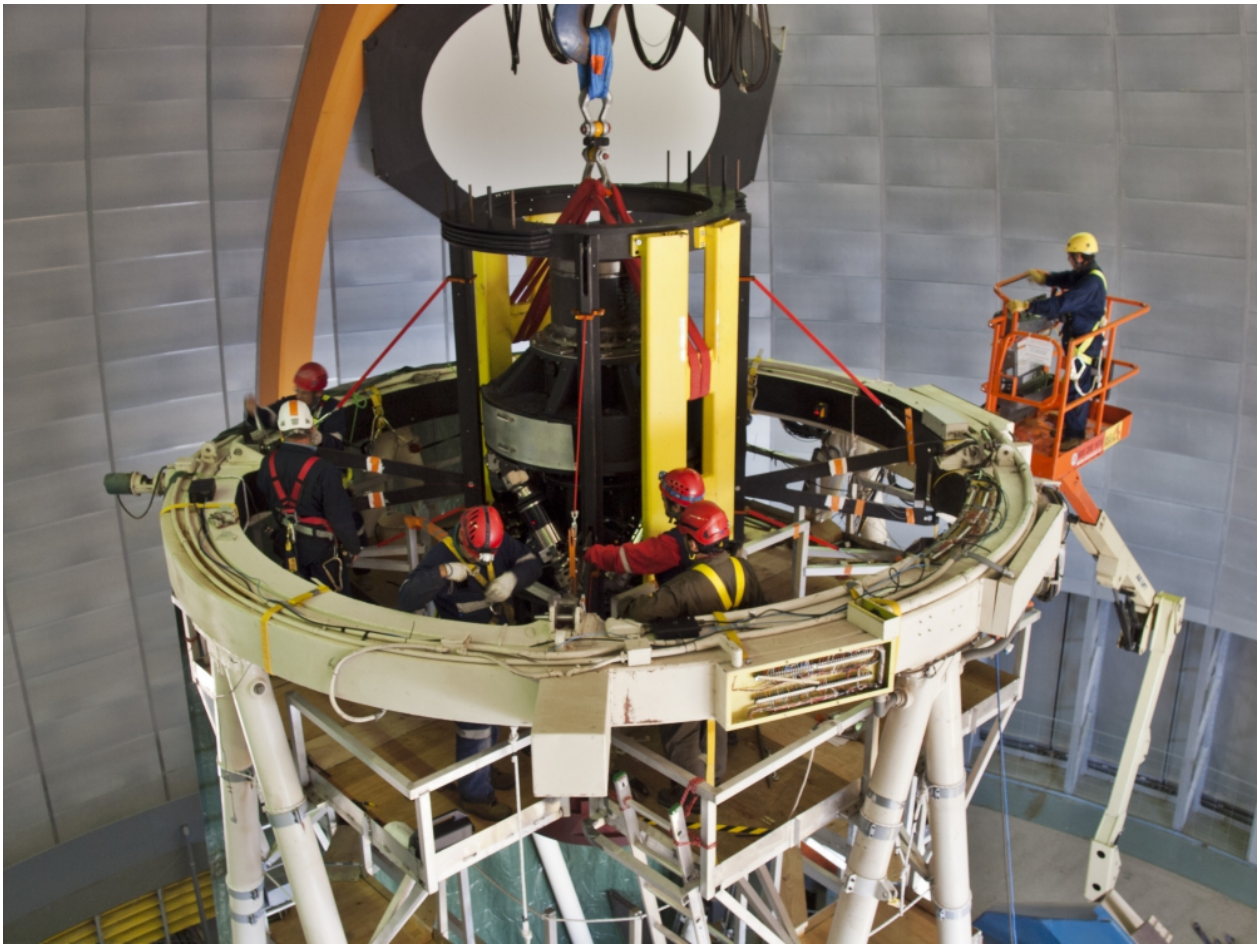

DECam installation on the Blanco Telescope:  cage, hexapods, corrector.  All of these components will be used for DESpec.  Photo credit Tim Abbott & NOAO/AURA/NSF.

# Executive Summary

We describe an initiative to build and use the Dark Energy Spectrometer (DESpec), a wide-field spectroscopic survey instrument for the Blanco 4-meter telescope at Cerro Tololo Inter-American Observatory (CTIO) in Chile. A new system with about 4000 robotically positioned optical fibers will be interchangeable with the CCD imager of the existing Dark Energy Camera (DECam), accessing a field of view of 3.8 square degrees in a single exposure. The proposed instrument will be operated by CTIO and available for use by the astronomy community. Our collaboration proposes to use DESpec to conduct a wide, deep spectroscopic survey to study Dark Energy, sharing the survey products with the community, while providing community access to fibers in parallel with survey observations.

In a survey of ~350 nights, the DESpec collaboration proposes to obtain spectroscopic redshifts for ~8 million galaxies over 5000 sq. deg. selected from precision imaging from the Dark Energy Survey (DES). This Dark Energy Spectroscopic Survey will advance our knowledge of cosmic expansion and structure growth significantly beyond that obtainable with imaging-only surveys. Since it adds a spectroscopic third dimension to the *same sky* as DES, the world's largest and deepest photometric survey, DESpec will enable a powerful array of new, increasingly precise techniques to discriminate among alternative explanations of cosmic acceleration, such as Dark Energy and Modified Gravity. In the 2020 decade, a wider-area spectroscopic survey will enhance the science reach of the next-generation imaging project, LSST, by providing spectroscopy for tens of millions of galaxies and QSO's over its 15,000-20,000 square degree survey area. DESpec will create a comprehensive, three dimensional, critically-sampled map of large scale cosmic structure – a database encompassing the "cosmic web" of galaxies and clusters over the entire, deeply-imaged southern sky – over most of cosmic history.

DESpec will take advantage of the substantial hardware infrastructure recently built for DECam to achieve excellent science at low cost and low technical and schedule risk. It will be mounted in the DECam prime focus cage with its hexapod positioning system (recently installed on the Blanco Telescope) and will share the four largest DECam optical corrector elements. It will be routinely interchangeable with the DECam CCD imager and preserve f/8 secondary capability, to allow flexible use of the telescope. The robotically positioned optical fibers will feed 10 relatively simple, high-throughput spectrometers outfitted with existing, spare, red-sensitive, science-grade DECam CCDs. An Atmospheric Dispersion Compensator similar to existing designs can be accommodated in the existing DECam filter slot. Examples that meet requirements for all major subsystems are in existing or highly developed instruments, showing that cost, technical, and schedule risks are manageable.

# 1. Introduction

## 1.A. The Foundational Impact of a Wide, Deep Spectroscopic Survey

In 1998, two teams of astronomers studying distant Type Ia supernovae presented evidence that the expansion of the Universe is speeding up rather than slowing down due to gravity (Riess, et al. 1998, Perlmutter, et al. 1999). In the following years, maps of cosmic structure produced by the Sloan Digital Sky Survey (SDSS), and deep maps of the cosmic microwave background by Boomerang, MAXIMA, DASI, WMAP and other experiments, confirmed and extended this result, consolidating for the first time a precise, comprehensive model for the overall size, shape, and contents of the universe (e.g., Tegmark, et al. 2004, Jaffe, et al. 2001, Pryke, et al. 2002, Spergel, et al. 2003). These transformational discoveries were recognized as *Science* magazine's "Breakthroughs of the Year" in 1998 and 2003. In 2011, Saul Perlmutter, Adam Riess, and Brian Schmidt were awarded the Nobel Prize in Physics for the discovery of cosmic acceleration.

While cosmology is now on a much more precise footing than before, the physical origin of cosmic acceleration remains a mystery (e.g., Frieman, Turner, & Huterer 2008; Sullivan, et al. 2011). Unraveling it will have profound implications for fundamental physics. Is acceleration caused by an exotic new form of Dark Energy (DE) that makes up most of the Universe? Or, does it indicate that Einstein's Theory of General Relativity (GR) must be replaced by a new theory of gravity on cosmic scales? If the answer is dark energy, is it the energy of the vacuum – Einstein's "cosmological constant" – or something else, perhaps an ultralight scalar field, dubbed quintessence? Some physicists even speculate that cosmic acceleration is a sign of new emergent physics that transcends our current separate notions of mass-energy and space-time.

This enduring and profound mystery can be addressed with better measurements of the history of the cosmic expansion rate and of the growth of large-scale structure. Ambitious next steps are already underway: the Dark Energy Survey (DES), starting later this year, and the Large Synoptic Survey Telescope (LSST) in the 2020 decade, will extend wide field imaging well beyond the depth of SDSS. They will create precision images of the sky with hundreds of millions to billions of galaxies, deep and broad enough to probe the history of cosmic acceleration and structure formation in unprecedented detail. Like SDSS, their databases of the deep sky will lead to a wide range of new probes and discoveries.

While the SDSS combined imaging and spectroscopy in a single survey, DES and LSST are imaging-only projects: they do not yet include a comprehensive spectroscopic survey. Detailed spectra, with resolution of several thousand, were important contributors to many discoveries with SDSS: they provide precision redshifts and distances and a wealth of information about the light-emitting sources as well as absorbing matter along the line of sight. Spectroscopic information will be even more critical in future studies, as we seek higher precision probes and better control over many unforeseen systematic errors.

In this document, we describe the Dark Energy Spectrometer, a low-risk, cost-effective way to create a wide, deep spectroscopic survey to accompany the DES and LSST imaging surveys. The system uses many existing components of the Dark Energy Camera, such as the wide-field corrector optics. The DECam CCD imager is removed and replaced by a new robotic fiber system that measures simultaneous spectra of about 4000 galaxies in a field of 3.8 square degrees.

This system critically samples *all* large-scale spatial structure in our past light cone out to a redshift of about 1.5, as traced by galaxies with mean separation ~15 $h^{-1}$Mpc. Their redshifts provide a high-resolution, comprehensive, three-dimensional map of the cosmic web extending over most of its history – from here and now back to before large-scale structure formed and before the cosmic expansion started to accelerate.

Using DES imaging to select targets, DESpec will cover about 5000 square degrees with this depth and sampling in just a few hundred nights. We show below how this large survey, with about 8 million spectra of DES sources, will provide new and uniquely powerful probes of dark energy physics.

This large initial survey could be in place at or near the beginning of the LSST survey. The precise map of large-scale structure would give a powerful boost to a wide range of early LSST science. After several more years, using LSST to select additional targets, the spectroscopic survey will be extended to the entire southern sky and over 20 million sources.

The rest of this white paper emphasizes the value of the DESpec survey for understanding the physics of cosmic acceleration in ways we can calculate today and provides initial concepts for the DESpec design. We note however that as with SDSS, this massive and versatile spectroscopic database will certainly amplify the science impact of DES and LSST over a much broader range of enquiry in many ways we cannot yet foresee, and that may indeed lead to even more profound discoveries.

## 1.B. DESpec: the Dark Energy Spectrometer and Spectroscopic Survey

1.B.1 The Dark Energy Survey

The Dark Energy Survey (DES) is a deep, wide, multi-band imaging survey, spanning 525 nights over six seasons beginning in late 2012, which will use the new 570-Megapixel Dark Energy Camera on the Blanco 4-m telescope at CTIO (DES Collaboration, 2005). DES, in partnership with ESO's near-infrared VISTA Hemisphere Survey (VHS), will provide imaging of ~300 million galaxies in 5+3 optical+NIR filters (grizY for DES, JHK for VHS) over 5000 sq. deg. DES will also discover and measure ~4000 SN Ia light curves in a time-domain survey of 30 sq. deg. With this survey, DES will probe dark energy using four techniques: the clustering of galaxies on large scales, including baryon acoustic oscillations (BAO); the abundance of massive galaxy clusters; weak gravitational lensing distortions of the images of distant galaxies; and Type Ia supernova distances. In the language of the Dark Energy Task Force report (DETF, Albrecht, et al. 2006), DES is a 'Stage III' experiment that will make a substantial step forward in constraining the properties of dark energy. DES is an international collaboration, with over 130 senior scientists from 27 institutions in the US, the UK, Spain, Brazil, Germany, and Switzerland.

1.B.2 Complementary Wide-Field Spectroscopic Survey

As a multi-band imaging survey, DES (and later LSST) will provide precise measurements of galaxy fluxes, colors, and shapes but only approximate photometric estimates of their redshifts (photo-z's). High-precision redshifts, which enable a true 3d map of the cosmos, require spectroscopy. The DESpec collaboration seeks to substantially enhance the science reach of DES imaging, and better probe the origin of cosmic acceleration, by obtaining spectroscopic redshifts for a large sample of DES target galaxies, yielding a dense sampling of 3d structure over the wide, deep volume probed by DES.

The need for such data motivates DESpec, a concept for a ~4000-fiber spectrograph for the Blanco telescope, which enables an ~8 million spectroscopic galaxy survey in ~350 nights. We consider DESpec an 'upgrade' of the Dark Energy Camera and of the DES project, following the model of upgrades that enhance the capabilities of high-energy particle physics experiments. Together with DES, DESpec has the science reach of a next-generation, DETF Stage IV project. Once LSST begins survey operations from neighboring Cerro Pachon, the DESpec survey can be expanded to ~15,000 sq. deg., yielding redshifts for ~23 million galaxies in ~1000 nights, substantially enhancing the science reach of LSST. This sample is comparable in size to, and complements, the proposed BigBOSS survey in the north (Schlegel, et al. 2011; see Appendix).

However, the unique strength of the DESpec survey transcends the statistics captured in the DETF figure of merit: it will provide a 3d redshift map of the Universe over the same deep, wide volume precisely mapped by DES and later LSST. The statistical versatility and power of a comprehensive, deep 3d survey coupled with a precision, multi-band, homogeneous photometric survey was demonstrated by SDSS, which was instrumental in establishing the current cosmological paradigm. To probe the physical origin of cosmic acceleration will require a combination of many statistical techniques, only some of which are now known and well tested. DESpec will survey luminous red galaxies to $z \sim 1$ and emission-line galaxies to $z \sim 1.5$, extending back to an era before Dark Energy dominated the cosmic expansion. DESpec enables a wide range of DE probes that powerfully synergize in ways that no other foreseen spectroscopic surveys can achieve.

1.B.3 Physics Reach

The spectroscopic redshift information that DESpec adds to the DES(+VHS) galaxy catalog results in a substantial increase in the precision of the dark energy equation of state parameter, *w,* and its time evolution, *dw/da,* from all four of the techniques above (baryon acoustic oscillations; abundance of massive galaxy clusters; weak gravitational lensing; Type Ia supernova distances). Spectroscopic redshift precision also enables qualitatively new dark energy probes beyond DES, for example, radial baryon acoustic oscillations and redshift-space distortions (RSD). It can provide dynamical mass estimates for thousands of DES galaxy clusters, strengthening the cluster probe of DE. Spectroscopic data also increase the power of weak lensing as a probe by controlling intrinsic alignment effects. In the southern hemisphere, this includes cross-correlation with lensing of the microwave background radiation, now detected in high-resolution imaging by the South Pole Telescope and the Atacama Cosmology Telescope (Das, et al. 2011, van Engelen, et al. 2012, Bleem, et al. 2012) .

Among the many new capabilities enabled by DESpec, one exciting recent realization is that the combination of redshift-space distortions from DESpec and weak lensing from DES (and later LSST) can powerfully discriminate models of Modified Gravity from Dark Energy as the cause of cosmic acceleration (Zhang, et al. 2007, Guzik, et al. 2009, Song & Dore 2009, Reyes, et al. 2010, Song, et al. 2011), *with reduced uncertainty due to galaxy bias and cosmic variance if the photometric and spectroscopic surveys cover the same sky area* as DES and DESpec would do (Pen 2004, McDonald & Seljak 2009, Bernstein & Cai 2011, Cai & Bernstein 2012, Gaztanaga, et al. 2012). Below we provide some further analysis of same-sky imaging and spectroscopy and outline our R&D program to study it further. Three independent studies indicate that same-sky for imaging and spectroscopy provides superior dark energy constraints to equivalent non-overlapping photometric and spectroscopic surveys.

Further, the DESpec survey can set a strict upper limit on or even detect the mass of neutrinos, with expected sensitivity of ~0.05 eV in combination with DES and Planck, highly competitive with laboratory experiments. This very large spectroscopic survey will also probe the presently puzzling excess power in the galaxy clustering power spectrum on the Gigaparsec scale (Thomas, Abdalla, & Lahav 2011) and enable new studies of galaxy evolution and of quasars. Additional payoffs include reducing DES weak lensing intrinsic alignment systematics by cross-correlating the lensing shear signal with a spectroscopic galaxy sample, improved determination of the redshift distribution of DES photometric galaxies via angular cross-correlation with a spectroscopic sample, thereby reducing systematic errors of all of the Dark Energy probes, detailed study of the dark matter environments of galaxies and clusters via stacked weak lensing mass estimates, and measurement of cluster dynamical masses via velocity dispersions. These arguments are strengthened when extended to wider-area follow-up of LSST imaging.

1.B.4 Access to the Southern Hemisphere

A variety of factors combine to argue for a deep, wide spectroscopic survey from the southern hemisphere. A well calibrated, uniform, deep target list of objects with precision multi-band photometry and lensing shape measurements is guaranteed from DES. DESpec can access the entire sky area that will be surveyed by both DES (5000 sq. deg.) and later LSST (20,000 sq. deg.), as well as the highest resolution mappers of the cosmic microwave background, the South Pole Telescope and the Atacama Cosmology Telescope. Located in the southern hemisphere at one of the world's premier astronomical sites (CTIO has a median site seeing of 0.65" FWHM and 80% useable nights), DESpec on the Blanco will yield a wealth of spectroscopic survey information complementary to the larger-aperture, narrower-field VLT, Gemini, Magellan, and other telescopes concentrated in the southern hemisphere. In the farther future, southern sites are currently planned for two of three next-generation very large optical telescopes, as well as the Square Kilometer Array in the radio. A similar wide-field spectroscopic survey currently under development, the BigBOSS project on the Kitt Peak Mayall 4-meter telescope, is in the northern hemisphere. The Sumire Prime Focus Spectrograph, under development for the larger-aperture Subaru Telescope, has a smaller field of view and is also in the northern hemisphere. ESO's 4MOST project has a wider field-of-view but is optimized for spectroscopic follow-up of stars astrometrically measured by Gaia.

In addition to its dark energy goals, as a community instrument DESpec will enable a wide array of coordinated spectroscopic surveys, affording opportunities for discoveries in stellar structure and evolution, nearby galaxies, galaxy evolution, the structure of galaxy clusters, and other applications. Given the large number of fibers, building on the experience of the Sloan Digital Sky Survey, one can optimize efficiency by conducting many survey programs in parallel. In particular, fibers will be available for parallel community observations during the DESpec survey. The ability to interchange the instrument with DECam enables a flexible observing program for the Blanco, including wide-field imaging in the pre-LSST era. This program efficiently exploits the unique wide field and structural capabilities of the Blanco telescope. The DESpec/DECam system creates an opportunity for optimal use of resources by both DOE and NSF user communities: a large science impact per dollar and a beneficial partnership combining the technical resources of both agencies. In the LSST era, the wide-field spectroscopic capability of DESpec complements two other capabilities – one following up very faint LSST sources over small sky areas (e.g., with ~8 to 30-meter telescopes and with JWST) and the other following up optical LSST transients on very rapid timescales – to form an optimized southern spectroscopic system.

1.B.5 Technical Concept

DESpec achieves relatively low cost, schedule, and technical risk by capitalizing on and leveraging the recent investment in DECam and recent structural improvements in the Blanco telescope and its environment made by NOAO (new primary mirror radial supports, improved telescope control system, environmental control, and other infrastructure). It uses the DECam mechanical structure (prime focus cage, barrel, shutter, hexapod alignment system), four of the same corrector lenses, and an ample supply (~60) of existing spare, packaged and tested, red-sensitive, science-grade 2kx4k CCDs produced for the DECam project. Two new optical corrector lenses will be needed, with the need for an Atmospheric Dispersion Compensator still under review. The resulting optical beam is nearly telecentric. The preliminary reference design described here entails ~4000 robotically positioned optical fibers of diameter 1.75" (100 microns) over the existing DECam 3.8 sq. deg. field of view, feeding 10 double-arm spectrographs. We are also investigating the science reach and cost trade-offs of a single-arm design for the spectrographs.

One low-cost spectrograph design builds upon that used for the Hobby-Eberly Telescope Dark Energy Experiment (HETDEX). The CCD readout electronics will be copied from the low-read-noise DECam design, again saving cost and reducing risk. The two-arm (dichroic) spectrograph would have a blue side wavelength range $480<\lambda<780$ nm, and the red side would cover $750<\lambda<1050$ nm. Over this range, the optical corrector produces an appropriately small spot size of 0.4-0.63" FWHM. These design numbers – fiber size, wavelength range, spectral resolution, and resulting spectrograph design – will be optimized in the R&D phase. Since NOAO expects to operate DECam for at least five years beyond the end of DES (i.e., to 2022), DESpec will be designed to be interchangeable with DECam. The design preserves the Blanco's capability to support an f/8 secondary, allowing flexible use of the telescope with other instruments at the Cassegrain focus.

An R&D program for DESpec now underway focuses on: (1) optimizing the optical and spectrograph design, balancing science requirements against cost and schedule and demonstrating feasibility; (2) establishing requirements for and demonstrating technological feasibility of the fiber-positioning system, likely in coordination with the group developing the Echidna fiber-positioning system whose prototype appears to meet our requirements; (3) building upon DECam expertise to complete the design of the CCD readout electronics and mechanical design; (4) optimization of the survey strategy (field overlaps, fiber collision and placement, galaxy target selection, etc.) using realistic simulated spectra and the inhomogeneous galaxy distribution; and (5) further studying the figure-of-merit benefits of same-sky photometry and spectroscopy. The R&D program will produce a final design and construction cost and schedule estimates.

The collaboration will seek project funding from the US agencies DOE and NSF (with major hardware funding to be proposed to the DOE), international funding agencies, and the participating institutions, following the successful model of the DES project. A number of DES and non-DES institutions will participate. There is substantial interest in the project from foreign institutions. The UK STFC recently awarded a £160k grant for DESpec R&D. The Australian Astronomical Observatory (AAO), a long-time world leader in astronomical fiber positioning systems and surveys, has indicated keen interest in taking a lead role in construction of the fiber positioning system. Seed funding for R&D has also been made available from the Kavli Institute for Cosmological Physics.

We will propose that the instrument be operated by the National Optical Astronomy Observatory (NOAO), which operates CTIO, under a time allocation agreement with the collaboration to be negotiated (as was done for DES, which was allocated 525 nights in exchange for DECam). However, it is important to note that NOAO has not yet issued nor indicated that they will issue an announcement of opportunity for a new prime-focus instrument for the Blanco Telescope.

Section 2 of this white paper outlines the science case for DESpec, focusing on probes of dark energy and tests of General Relativity. Section 3 discusses the selection of spectroscopic targets and defines a strawman survey strategy that follows from the science requirements of Section 2. Section 4 discusses the DESpec instrument components, highlighting those areas that would be addressed in the R&D phase. An appendix briefly compares DESpec with the proposed BigBOSS project.

## 2. DESpec Dark Energy Science Program

### 2.A. Probing the origin of cosmic acceleration and testing General Relativity

The Dark Energy Survey (DES) will enable measurements of the dark energy and dark matter densities and of the dark energy equation of state through four methods: galaxy clusters, weak gravitational lensing (WL), galaxy angular clustering (including angular Baryon Acoustic Oscillations, BAO), and supernovae (SNe). A spectroscopic redshift survey of a substantial fraction of DES target galaxies with an instrument such as DESpec would enhance each of these techniques, particularly the measurement of baryon acoustic oscillations (BAO) through galaxy clustering, and enable a new DE probe, redshift-space distortions (RSD). Moreover, the combination of RSD from DESpec with WL from DES and subsequently LSST would enable a powerful new test of the consistency of the General Relativity-plus-Dark Energy paradigm and therefore help distinguish Dark Energy from Modified Gravity as the physical cause of cosmic acceleration.

To frame the discussion, we outline the DES survey and a strawman concept for the DESpec survey – the latter discussed in Sec. 3 – and introduce figures of merit for DE and parameters for testing departures from General Relativity.

The DES will comprise two multi-band imaging surveys, a wide-field survey and a narrow time-domain survey. The wide-field survey will nominally cover 5000 sq. deg. in the south Galactic cap, all at high galactic latitude suitable for extragalactic studies, reaching ~24$^{th}$ magnitude in the grizY filters. The depth and filter coverage of the wide-field survey were chosen primarily to achieve accurate galaxy and cluster photo-z and shape measurements to redshifts $z>1$. The wide-field survey will detect over 100,000 galaxy clusters and measure shapes, photo-z's, and positions for ~200 million galaxies. It will overlap completely with the ESO Vista Hemisphere Survey (VHS), which will obtain moderately deep imaging in J, H, and K filters. The combined 8-filter data will extend the range of precise galaxy photo-z's to $z$~2. The DES Supernova Survey involves frequent (every few days) imaging of a 30 sq. deg. area in the griz filters, which will yield well-measured light curves for ~4000 Type Ia supernovae to redshifts $z$~1.

The DETF defined a figure of merit (FoM) for dark energy surveys by parametrizing the redshift evolution of the dark energy equation of state parameter by $w(a)=w_0+w_a(1-a)$, where $a(t)=1/(1+z)$ is the cosmic scale factor, $w_0$ is the current value of $w$, and $w_a$ is a measure of its evolution with redshift. The DETF FoM is proportional to the reciprocal of the area in the $w_0$-$w_a$ plane that encloses the 95% CL region. Defining a pivot epoch, $a_p$, at which the uncertainty in $w(a)$ is minimized for a given experiment, the DETF FoM is $[\sigma(w_p)\sigma(w_a)]^{-1}$. More complex figures of merit have also been proposed (Albrecht & Bernstein 2007, Albrecht, et al. 2009). The DETF report provided an estimate for the Stage II FoM of about 60, where Stage II includes projections from surveys that were on-going at the time the report was written, combined with the forecast statistical precision of Planck CMB measurements on cosmological parameters. Using similar techniques, the DES collaboration estimated a combined FoM from all four techniques of about 260 for the final survey, characteristic of a DETF Stage III project. There is considerable uncertainty in this forecast, since there are large uncertainties in the ultimate levels of systematic errors for each of the techniques. According to the DETF, next-generation, Stage IV projects would be anticipated to increase the DETF FoM by another factor of ~3-5 compared to Stage III. Our initial projections indicate that the DES+DESpec combination would reach that level of precision.

In testing Modified Gravity (MG) vs. Dark Energy, it is also useful to have parameters and FoMs describing departures from General Relativity, and several have been proposed. In GR+DE, the linear growth rate of density perturbations $\delta(a)$ that form large-scale structure is uniquely determined by the expansion rate $H(a)$ and the matter density parameter $\Omega_m$. In particular, the logarithmic growth rate is given by $f(a)=d\ln\delta/d\ln a=\Omega_m(a)^\gamma$, where the growth exponent $\gamma=0.55$ in GR. In Modified Gravity theories, the relation between expansion history and perturbation growth can be changed; e.g., in the DGP braneworld model (Dvali, et al. 2000), $\gamma=0.68$. One FoM for modified gravity models is therefore $[\sigma(\gamma)]^{-2}$ (Albrecht, et al. 2009). More general MG parameterizations have also been considered, as discussed below in Sec. 3.

In addition to dark energy and modified gravity, DES+DESpec would probe other areas of fundamental physics, including neutrinos and primordial non-Gaussianity from inflation. Current constraints on the sum of neutrino masses from large-scale structure depend somewhat on details and assumptions of the analyses, but a relatively conservative recent analysis finds an upper bound of 0.28 eV at 95% CL (Thomas, Abdalla, and Lahav 2010), and DES+Planck is expected to reach 0.1 eV (Lahav, et al. 2010). We estimate that DES+DESpec clustering measurements plus Planck would improve this bound to ~0.05 eV, reaching the regime where neutrino oscillation experiments indicate a detection is likely. Measurement of large-scale clustering in DES+DESpec, particularly constraints on the scale-dependence of galaxy bias, will constrain departures from primordial Gaussianity and thereby test models of primordial inflation.

For our baseline DESpec survey, we assume ~8 million successful redshifts are acquired over the 5000 sq. deg. DES footprint. By using flux, colors, and surface brightness to target a mixture of Luminous Red Galaxies (LRGs) at $z<1$ and Emission Line Galaxies (ELGs) at redshifts $0.6<z<1.7$, we assume for purposes of illustration that the redshift distribution can be sculpted to be approximately constant over the redshift range $0.2<z<1.5$. Sec. 3 describes how such a selection could be carried out. An extended DESpec survey would extend this targeting to ~23 million galaxies over 15,000 sq. deg. of extragalactic sky, by selecting targets from LSST. We note that, once it is operational, LSST will immediately reach the depth needed for selecting DESpec targets over its full survey area. We emphasize that this survey plan is just a strawman; the final target selection for DESpec (redshift distribution, flux limits, color selection, mix of LRG and ELG targets, exposure times) will follow from detailed R&D and science trade studies.

## 2.B Weak Lensing and Redshift Space Distortions

DESpec spectroscopy would enable measurement of galaxy clustering in redshift space. A galaxy's redshift is the combination of its Hubble flow motion and the radial component of its peculiar velocity due to nearby structures. Clustering in redshift space is therefore distorted (anisotropic) relative to clustering in real space, due to the effects of peculiar velocities. In linear perturbation theory and assuming linear bias between galaxies and dark matter, the galaxy density perturbation Fourier amplitude in redshift space is given by (Kaiser 1987)

$$\delta_g^s(\mathbf{k}) = (b + f\mu^2)\delta(\mathbf{k})$$

where $b$ is the linear bias factor for the given population of galaxies, $f$ is the linear growth factor defined above, $\mu$ is the cosine of the angle between $\mathbf{k}$ and the line of sight (*not* the same $\mu$ as in Sec. 2.A), and $\delta(\mathbf{k})$ is the real-space dark matter perturbation Fourier amplitude. Measurement of the anisotropy ($\mu$-dependence) of the galaxy power spectrum in redshift space thus provides a measure of the growth rate of fluctuations, *f(a),* which in turn is sensitive to the properties of dark energy or modified gravity.

Figure 2.1 (from Gaztanaga, et al. 2011) shows a first estimate of the 68% CL statistical constraints on $w_0$, $w_a$, and $\gamma$ that would be expected with a baseline 5000 sq. deg. DESpec survey from redshift space distortions alone (RSD, purple), from DES weak lensing alone (Shear, blue), from the combination of RSD and weak lensing in non-coincident parts of the sky (red, RSD+Shear, e.g., from BigBOSS+DES assuming the baseline 5000 sq. deg. footprint of DES), and from the coincident combination of RSD+WL (yellow, i.e., DESpec+DES). It is found that the combination of weak lensing and RSD measurements in the same spatial volume (yellow), as DESpec plus DES would provide, leads to stronger constraints on dark energy and modified gravity (Bernstein & Cai, 2011, Cai & Bernstein 2012, Gaztanaga, et al. 2012).

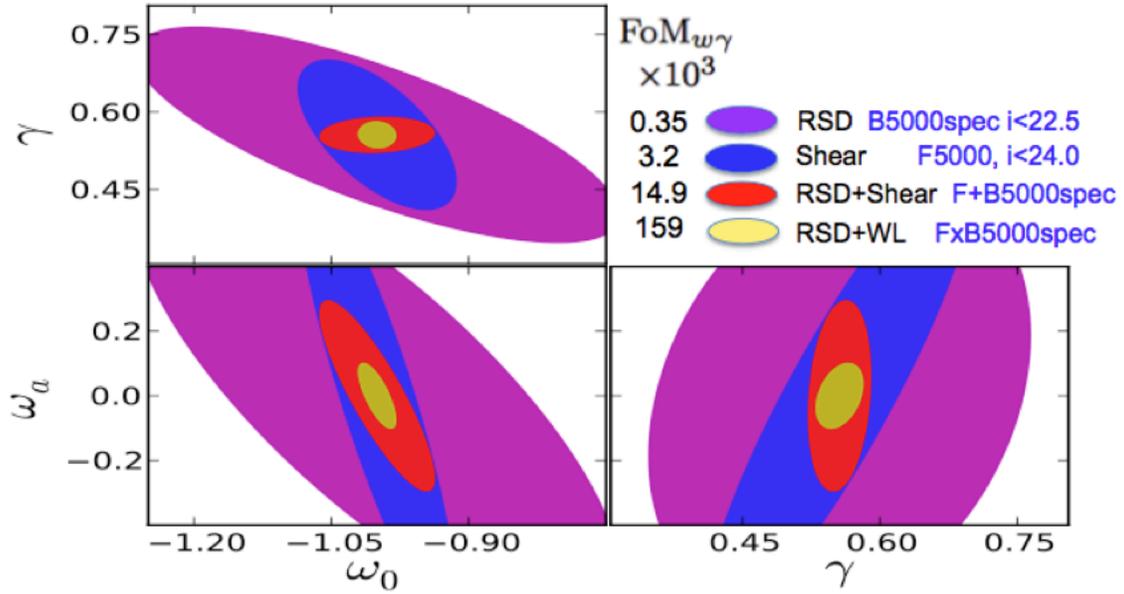

***Figure 2.1:*** *Forecast 68% CL constraints on dark energy ($w_0$, $w_a$) and structure growth ($\gamma$) parameters from DES weak lensing alone (blue), redshift space distortions alone (purple), the combination of the two without spatial overlap (red) and with spatial overlap (yellow, i.e., DES+DESpec). In the last case, the lensing measurements include shear-shear, galaxy-shear, and galaxy-magnification correlations. Each two-parameter combination is marginalized over the third parameter. We use Planck and Stage-II priors. From Gaztanaga, et al. (2011).*

In this calculation, the DESpec survey is assumed to obtain redshifts for 1000 galaxies per sq. deg. to limiting magnitude of $i_{AB}$=22.5, with a mean redshift of about 0.6. RSD×WL (DES+DESpec) improves the (marginalized over $\gamma$) DETF FoM for $w_0$, $w_a$ by a factor of about 4.6 compared to DES weak lensing alone. The spatial overlap of RSD spectroscopy with WL photometry is found to improve the DETF FoM compared to non-overlapping surveys. The improvement is around a factor 5-10 when one combines weak lensing and galaxy clustering (see Gaztanaga, et al. 2011), but note that this target selection choice does not optimize the survey strategy. Figure 2.2 shows which redshift bins are most important for the constraints, which in turn informs the optimization of target selection. Later (Section 3), we will consider more elaborate methods of selecting galaxies to demonstrate how the FoM can be further enhanced. We will also show how we reach similar conclusions with different assumptions and observables.

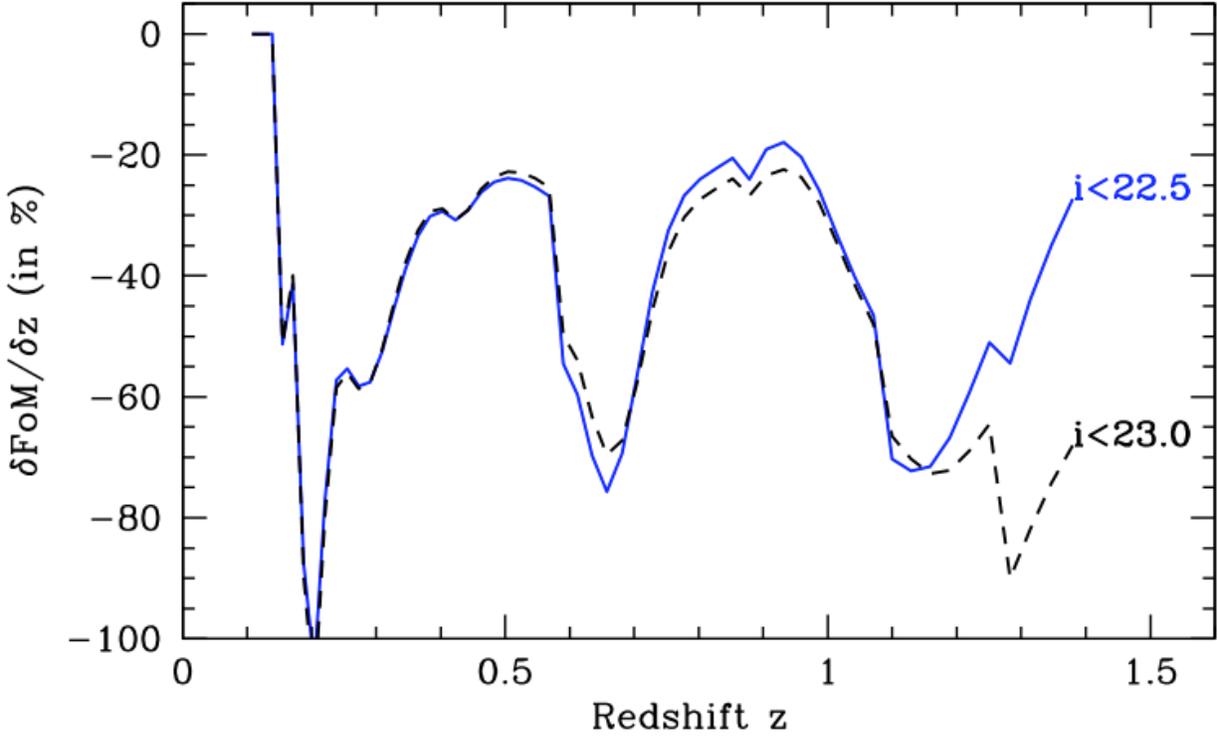

*Figure 2.2:* *Percentage change in joint constraint on dark energy ($w_0$, $w_a$) and structure growth ($\gamma$) for the galaxy shear +RSD (DES+DESpec) combination when we remove information from a particular redshift bin (relative change per unit redshift) for a DESpec survey limited to $i_{AB} <23.0$ (black dashed curve) and $i_{AB}<22.5$ (solid blue). The largest integrated contribution to this FoM comes from the highest redshifts, which sample the largest volumes. However, low and intermediate redshifts provide contributions of similar importance, due to the weak lensing efficiency and the measurement of bias, which is more optimal at low redshift. We can use this type of analysis to optimize DESpec target selection (see Section 3).*

A similar forecast has been done by Cai & Bernstein (2012), who focused on the growth parameter $\gamma$. Fig. 2.3 shows their comparison of the modified gravity FoM for surveys combining WL and RSD in separate volumes (non-overlapping) with those in the same volume (overlapping sky). For DES, we expect an effective lensing source density of $N_{lens}=10$ galaxies/arcmin$^2$; in this regime, Fig. 2.3 shows a gain of a factor of about 1.5-1.8 in the modified gravity FoM compared to non-overlapping surveys if the spectroscopic survey targets of order one galaxy per halo in all halos down to $M_{min} \sim 10^{13}$ $M_{sun}$. In the redshift range $z\sim 0.5$-1, this corresponds to a spatial target density of about $10^{-4}$ (h$^{-1}$ Mpc)$^{-3}$ or an areal density of ~300 per sq. deg. Luminous Red Galaxies (LRGs) would be ideal targets for this population. The same-sky enhancement in the growth parameter precision is more modest than the gains in the dark energy parameters but still significant.

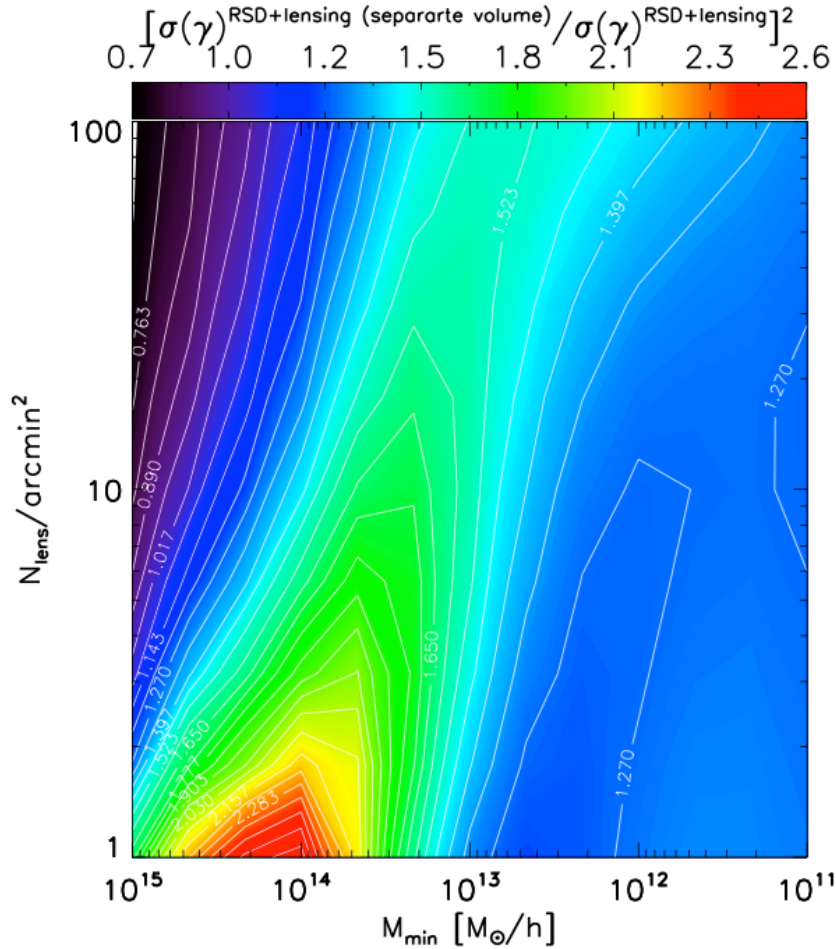

*Figure 2.3:* Inverse ratio of RSD+WL modified gravity Figure of Merit for non-overlapping vs. overlapping spectroscopic+photometric surveys. The design strategies for the DES and DESpec surveys correspond to a gain of a factor 1.5-1.8 in the MG FoM for overlapping vs. non-overlapping surveys. From Cai & Bernstein (2012).

## 2.C Large-scale Structure and Baryon Acoustic Oscillations

The large-scale clustering of galaxies contains a feature at ~110 $h^{-1}$ Mpc, detectable as a slight enhancement in the two-point correlation function on that scale or as periodic wiggles in the power spectrum. This scale is set by the sound horizon, the distance that acoustic waves in the photon-baryon plasma travelled by the time that ionized hydrogen recombined 380,000 years after the Big Bang. This BAO feature is imprinted dramatically on the cosmic microwave background anisotropy as a characteristic angular scale of ~1 degree in the size of hot and cold spots in the CMB temperature map. It is a more subtle feature in the galaxy distribution, since structure formation is driven primarily by gravitational instability of dark matter. First detected in the distribution of LRGs in the SDSS redshift survey (Eisenstein, et al. 2005), its measurement is the primary aim of the on-going SDSS-III Baryon Oscillation Spectroscopic Survey (BOSS), which is targeting 1.5 million LRGs over 10,000 sq. deg. to redshift $z\sim0.6$ (Eisenstein, et al. 2011, Dawson et al 2012, Anderson et al. 2012). A number of next-generation BAO-centric redshift surveys are under construction (e.g., HETDEX) or have been proposed (e.g., BigBOSS, WEAVE, VXMS, SUMIRE, EUCLID, WFIRST). The BAO feature is useful for dark energy studies, because it acts as a standard ruler (e.g., Blake & Glazebrook 2003, Seo & Eisenstein 2003). Photometric surveys such as DES and LSST will probe the BAO feature in angular galaxy clustering, measuring the angular diameter distance $D_A(z)$ and thereby constraining the expansion history. Redshift surveys improve the accuracy on $D_A(z)$ and, in addition, measure BAO in the radial direction, which provides a direct measure of *H(z)* (Gaztanaga, Cabre, & Hui 2009, Kazin, et al 2010). Greater precision in these geometric measures over a range of redshifts translates into greater precision in the dark energy equation of state.

The SDSS LRG BAO measurement constrained *H(z)* and $D_A(z)$ to ~4% and 1.6% precision at redshift $z\sim0.35$. BOSS is expected to determine these over a range of redshifts up to $z=0.6$, with its best precision of ~2 and ~1% at the upper end of that range. DES will probe $D_A(z)$ via angular BAO with a precision of about 2.5% but over a larger redshift range, to $z\sim1.3$. A DESpec survey that targets Emission Line Galaxies to $z\sim1.5$ over 5000 sq. deg. would reach a statistical precision of ~2.5% in *H(z)* and 1.5% in $D_A(z)$ at $z\sim0.9$ but with useful measurements over the entire redshift range. Extending the DESpec ELG survey to 15,000 sq. deg. by targeting LSST galaxies in addition would improve the precision on *H* and $D_A$ to about 1.5 and better than 1%; this precision would be achieved by BigBOSS as well. Combining DESpec and BigBOSS BAO would be expected to further improve each of these constraints by ~40% and give an independent check on systematic errors.

In designing a BAO survey, the precision of the *H* and $D_A$ measurements (and therefore of the DE constraints) is set by the relative statistical uncertainty in measuring the large-scale galaxy power spectrum, $\sigma(P)/P \propto V_{eff}^{-1/2}$, where $V_{eff}$ is the effective volume of the survey,

$$V_{eff} = \left(\frac{nP}{nP+1}\right)^2 V_{survey} ,$$

$V_{survey}$ is the survey volume, and *n* is the survey galaxy number density. To minimize cosmic variance errors, the survey should cover large volume $V_{survey}$; to minimize the impact of Poisson errors, the galaxy sampling density should satisfy $nP \geq 2-3$. To reach large effective volume, we wish to select galaxies with adequate sampling density out to high redshifts, $z\sim1.5$, which leads one to preferentially target Emission Line Galaxies (ELGs), as described in Sec. 3. Given the ELG large-scale power spectrum amplitude, the desired number density is very roughly $n > 3\times10^{-4}$ $h^3$ Mpc$^{-3}$; for a 5000 sq. deg. survey to that redshift, the survey volume is of order $2\times10^{10}$ $h^{-3}$ Mpc$^3$, which implies an approximate total of $N=nV_{survey}\sim6$ million ELG redshifts, with an areal density of 1200 per sq. deg. This figure sets the survey design of Sec. 3. For a wider-area BAO survey that also targets LSST galaxies, one would seek to maintain the same target density, resulting in ~18 million ELG redshifts over 15,000 sq. deg.

## 2.D Galaxy Clusters

DES will measure the abundance of massive galaxy clusters to redshift $z$~1.3 and use it to probe DE. The abundance of dark matter halos with mass and redshift, $dN(M,z)/dzd\Omega$, is sensitive to DE through the expansion history, which determines the volume element, and to the growth rate of structure, which determines the intrinsic halo density as a function of time. In this technique, clusters serve as proxies for massive dark matter halos. DES will determine cluster redshifts photometrically with sufficient precision for this test. However, since the masses of dark matter halos are not directly observable, this technique relies critically on the ability to determine the relation between cluster observables (such as the number of red galaxies above a certain luminosity) and the underlying halo masses; in particular, the mean and scatter of the mass-observable relation must be determined either internally (e.g., via self-calibration or weak lensing) or externally, by performing other measurements that constrain the relation. Clearly, external measurements that help reduce the uncertainty in the cluster mass-observable relation will strengthen the resulting DE constraints (Cunha 2009, Wu, Rozo, & Wechsler 2010). DESpec can provide such external information for DES clusters by providing dynamical cluster mass estimates: measuring redshifts for several tens of galaxies in a cluster provides an estimate of its velocity dispersion and therefore of its mass. Well-designed spectroscopic follow-up could improve the DES cluster DE FOM by a factor of several even if carried out over a subset of DES clusters. Targeting for this sample would synergize with the LRG targeting for RSD and BAO studies but would likely require additional cluster-optimized exposures as well.

## 2.E Supernovae

The DES Supernova Survey will measure high-quality light curves for ~4000 type Ia supernovae to redshifts $z$~1 through repeat imaging of a 30 sq. deg. region over the course of the survey. Given limited spectroscopic resources, only a fraction of those will have follow-up spectroscopy while the supernovae are bright enough to detect. For the rest, follow-up spectroscopy of the host galaxies will be needed to precisely determine the SN redshift, which is important for increasing the precision of the SN Hubble diagram and reducing contamination from non-Ia SN types. The spectroscopic measurement of SN Ia host-galaxy properties (star-formation rates, gas-phase metallicities, etc) has become even more important with the results of recent studies (Kelly, et al. 2010, Lampeitl, et al. 2010, Sullivan, et al. 2010, Gupta, et al. 2011, D'Andrea, et al. 2011) indicating that SN Ia luminosities (and therefore distance estimates) are correlated with host-galaxy properties. This correlation must now be taken into account in order to control SN cosmology systematic errors due to evolution of the mix of host-galaxy types (Sullivan, et al. 2011).

DESpec could contribute significantly to this program by measuring host-galaxy spectra of a large fraction of the DES SN Ia sample. Different science goals can be achieved, depending on the depth of the spectroscopic observations. At the lowest S/N ratio of ~3-5 in the continuum, we can measure accurate redshifts and star-formation rates of most emission-line host galaxies. Approximately 30% of the hosts in this category are brighter than r=22 mag, which can be achieved with DESpec in 30-minute exposures. A more ambitious observing program can measure redshifts of emission-line galaxies as faint as r=24 mag with total integration times of several hours. This will provide redshifts of ~90% of all SN Ia host galaxies from DES. The SNLS group, for example, has demonstrated the efficient use of the 4m Anglo-Australian Telescope and the AAOmega multi-object spectrograph for measuring SN Ia host galaxy redshifts down to r=24 mag with 60 ksec exposures (Lidman, et al. 2012).

At S/N of ~10 in the continuum, we can measure the gas-phase metallicities of the hosts, which is an important parameter that is correlated with the Hubble residual at z<0.15 (D'Andrea, et al. 2011). DESpec can study this correlation at z<0.4 (r=22 mag limit) and z<0.7 (r=24 mag limit) and will allow us to check for possible evolution. We can also measure average stellar population ages and metallicities with the use of Lick indices (Burstein, et al. 1984), which requires much higher S/N of at least ~30-50. This can be achieved at the lowest redshifts (z<0.2) or by stacking the spectra in several SN Ia luminosity bins.

## 2.F Photometric Redshift Calibration

All four DE probes in DES rely on accurate estimation of galaxy photometric redshifts using color and other information from the DECam grizY images. The DES survey strategy is designed to optimize galaxy photo-z measurements to redshifts $z>1$, and the synergy of DES with the near-infrared (JHK) ESO Vista Hemisphere Survey will improve the photo-z precision of the survey. The weak lensing and galaxy clustering measurements in DES in particular rely on accurate estimation of the galaxy redshift distribution, *N(z)*, and therefore require accurate estimation of the photo-*z* error distribution as a function of redshift. Uncertainties in the variance and bias of photo-z estimates lead to systematic errors that degrade DE constraints (Huterer, et al. 2005, Ma, Hu, & Huterer 2006).

Determining the error distribution as well as training empirical photo-*z* estimators requires spectroscopic samples of $\sim 10^4$-$10^5$ galaxies that cover the range of galaxy properties (colors, redshifts, flux limits, etc) of the photometric sample. While such spectroscopic samples in the DES survey footprint exist to approximately the DES depth, they are incomplete at the faintest magnitudes DES will reach.

The DESpec survey could aid in DES photometric redshift calibration in several ways. First, it will provide a large number of additional redshifts for training, validation, and error estimation for empirical photo-z methods, although not to the flux limit of the DES. Second, such a large sample of galaxy redshifts overlapping the DES footprint can in principle enable new methods that can augment traditional photo-z estimates at faint magnitudes. In particular, since galaxies are clustered, the angular cross-correlation between overlapping spectroscopic and photometric samples spanning the same redshift range can be used to estimate the redshift distribution of the photometric sample, calibrating the true redshift distribution for DES galaxies in photo-z slices (Newman 2008, Matthews & Newman 2010). Reducing the uncertainty in $N(z)$ leads directly to improved dark energy constraints from the photometric sample. Initial estimates indicate substantial improvement in DES photo-z calibration from the technique for a spectroscopic survey covering ~60% of the DES area. Third, for a spectroscopic survey that completely overlaps DES, one can potentially take advantage of the fact that a faint galaxy in the DES photometric survey that is near on the sky to a brighter galaxy with a DESpec redshift will have a reasonable probability of having the same redshift as the spectroscopic galaxy. The DESpec redshifts therefore act as informative redshift priors for neighboring DES galaxies (Kovac, et al. 2009); the utility of this approach for DE photometric surveys is under study.

## 2.G Galaxy Evolution

While cold dark matter cosmologies provide a powerful framework to describe the formation and evolution of galaxies, we still have a poor understanding of key aspects of galaxy evolution including gas accretion and star formation processes, the mechanism for star formation quenching, relative roles of major and minor galaxy mergers, galaxy mass growth, and the role of the environment. Because of the complexity of these physical processes driving galaxy formation, progress in observational studies of galaxies can only be made through a comprehensive mapping of galaxy properties over a large range of time. DESpec will deliver this. The DESpec spectroscopic data set will create an enormous legacy value for a large range of additional science including studies of galaxy formation and evolution. Spectra of over one million luminous red galaxies out to redshifts z~1 will leverage the exploitation of the DES imaging data set and will lift the currently ongoing galaxy evolution studies of SDSS-III/BOSS data to a new level by pushing to larger look-back times.

DESpec will allow us to analyze stellar population properties, chemical enrichment histories, gas physics, dark matter content, structural properties, galaxy mass functions, merger rates, and number densities for a variety of galaxy types as a function of environment and look-back time to ultimately constrain the formation and assembly histories of galaxies. The large volume of the DESpec observations will allow the exploitation of a large and diverse parameter space of galaxy environment, type, color, luminosity, mass and size. The accurate spectroscopic redshifts produced by DESpec will boost the analysis of the DES imaging data set significantly. The legacy value of DESpec spectroscopy will go well beyond the use of spectroscopic redshifts. The typical signal-to-noise ratio will be sufficient to make measurements of fundamental galaxy properties such as stellar and gas kinematics. The true power, however, will lie in the possibility of stacking hundreds and thousands of spectra to produce a spectroscopic data set of galaxies up to z ~ 1 with data quality unmatched by any of the existing spectroscopic galaxy surveys. From stacked

DESpec spectra we can derive accurate stellar masses, ages, and element abundances ratios by fitting state-of-the-art stellar population models (Thomas, et al. 2011; Maraston & Strömbäck 2011), which allows us to tap into the fossil record of galaxies and derive accurate formation and chemical enrichment histories (Thomas, et al. 2005, 2010). DESpec will clearly open a new chapter in observational galaxy evolution.

## 2.H Opportunities for Community Science

The SDSS demonstrated the extraordinary value of combining photometry with spectroscopy, where high-precision photometry is used to select targets for spectroscopy. The SDSS was designed around a focused set of scientific requirements (large-scale structure in the galaxy distribution, and quasars), yet its legacy data products have became one of the most widely used astronomical archives ever created, leading to thousands of publications spanning from the Solar System to cosmology. One can confidently predict that the photometric surveys in the southern hemisphere (DES, VISTA, LSST, and others), in concert with DESpec spectroscopic targeting, will have at least comparable impact for the world astronomical community.

DESpec will be a facility instrument at CTIO, providing open access to wide-field spectroscopy in the southern hemisphere. The southern hemisphere features unique astrophysical environments in the Galactic center, the Magellanic Clouds, and star-forming regions elsewhere in the southern Milky Way. Importantly, DESpec will see the same sky as LSST, which will produce a deep, high-precision, time-resolved map of the southern sky. The source catalog generated by LSST will have unprecedented value, and DESpec's capability to follow up sources discovered and characterized by LSST will yield enormous scientific leverage. Put differently, spectroscopic follow-up is required to fully exploit LSST imaging, and DESpec on the Blanco Telescope responds to that need.

The DESpec reference design can be modified or enhanced to enable more community science – we are open to community input about design priorities. For example, the fiber-positioning technology we will pursue allows additional fibers to be placed in the focal plane. These additional fibers could observe "community targets" in parallel with the DESpec cosmological survey. There are a number of other options for community access, including standard PI-directed use of the instrument where all the targets are selected for a specific observing program. Alternatively, one could combine several target classes from different programs, each with similar requirements on exposure time. In effect, several surveys can run in parallel: since DESpec can observe 1000 targets per square degree, this mode is equivalent (say) to having 10 SDSS-like spectroscopic surveys packaged together, each with 100 targets per square degree.

An example of a potential DESpec community program is follow-up spectroscopy of stars measured by Gaia for proper motion (an idea that motivates ESO's proposed 4MOST multiple-object spectrometer). Gaia will produce an astrometric catalog to $V = 20$, faint enough to guarantee that all of the DESpec fibers are subscribed. DESpec stellar radial velocities will be accurate to better than 4 km/sec (based on SDSS), and stellar abundances can be determined to better than 0.25 dex, again using SDSS experience. Such a survey would densely sample the position-velocity-abundance phase space for millions of stars in volumes not accessible in the northern hemisphere, providing basic information about the history of the stellar build-up of the Milky Way.

## 3. Survey Definition and Requirements

The scientific goals of the DESpec spectroscopic survey lead to a survey definition that acquires at least 1500 (=1300 ELG+230 LRG) galaxy redshifts per square degree up to $z$~1.5 over at least the DES footprint of 5000 square degrees, or a total of 7.6 million successful redshifts. The area will be extended up to 15,000 square degrees by later using LSST photometry for spectroscopic target selection. The DES and LSST photometry (together with VISTA JHK photometry) yields not only fluxes, colors, and photometric redshifts but also galaxy image shapes and surface brightnesses. All of this information can be exploited to select a sample of galaxies that satisfy the joint requirements of large redshift range, adequate volume sampling, and control over any bias introduced due to sample selection or redshift failures. In practice, we expect to use galaxy flux, color (and photo-z), and surface-brightness to optimize the redshift distribution and galaxy types of the survey. We plan to target a mix of emission-line galaxies (ELGs) – which predominate at high redshift and which yield efficient redshift estimates to $z$~1.7 based upon their prominent emission lines – and luminous red galaxies (LRGs), which have brighter continuum spectra and higher clustering amplitude and offer good redshift success rates up to $z$~1. The target selection will be chosen to optimize the science yield; at this stage, we present some examples of survey selection to demonstrate the feasibility of the survey, to set the scale, and to derive constraints upon the instrument design. A key question is whether we need to maximize high-redshift galaxies (z>1) in contrast to a better sampling of the whole galaxy population at intermediate redshifts (z~0.5-1.0); the latter covers less volume but optimally complements weak lensing science from DES (see Fig. 2.2). The combination of DES and DESpec can be used to measure galaxy biasing and recover the full modes of the 3d matter distribution, as opposed to just BAO or RSD information. The DESpec sample optimization is being worked out in detail, but we next show some examples that illustrate the potential of combining DES with DESpec.

### 3.A Simulated Galaxy Catalogs

We have created simulated catalogs of galaxies to quantify the yield of particular spectroscopic selection schemes, to predict the number of potential targets, and to compute the sensitivities and thereby the rate of progress of the survey. The simulations are built directly from the observed COSMOS catalog of Capak, et al. (2009) and Ilbert, et al. (2009). We refer to this simulation as the COSMOS Mock Catalog (CMC) (Jouvel, et al. 2009).

The COSMOS photometric-redshift catalog (Ilbert, et al. 2009) was computed with 30 bands over 2 deg$^2$ (GALEX for the UV bands, Subaru for the optical (U to z), and CFHT, UKIRT, and Spitzer for the NIR bands). It achieves very good photo-z accuracy and low catastrophic redshift rates due to careful calibration with the spectroscopic samples zCOSMOS and MIPS. The CMC is restricted to the area fully covered by HST/ACS imaging, 1.24 deg$^2$ after removal of masked areas. There are a total of 538,000 simulated galaxies for i < 26.5 in the mock catalog, leading to a density of roughly 120 gal/arcmin$^2$. AGN, stars, and X-ray sources were removed from the input COSMOS catalog.

A photo-z and a best-fit template spectrum (including possible additional extinction) are assigned to each galaxy of the COSMOS mock catalog. We first integrate the best-fit template through the instrument filter transmission curves to produce simulated magnitudes in the instrument filter set. We then apply random errors to the simulated magnitudes based upon a simple magnitude-error relation in each filter. The simulated mix of galaxy populations is then, by construction, representative of a real galaxy survey, and additional quantities measured in COSMOS (such as galaxy size, UV luminosity, morphology, stellar masses, correlation in position) can be easily propagated to the simulated catalog. The COSMOS mock catalog is limited to the range of magnitude space where the COSMOS imaging is complete ($i_{AB}$ ~ 26.2 for a 5 sigma detection, see Capak, et al. (2007) and Capak, et al. (2009)).

We assign emission-line fluxes to each galaxy of the CMC by modeling the emission-line fluxes (Ly alpha, [OII], H beta, [OIII], and H alpha) using the Kennicutt (1998) calibration. We first estimate the star-formation rate (SFR) from the dust-corrected UV rest-frame luminosity already measured for each COSMOS galaxy. The SFR can then be translated to an [OII] emission-line flux using another calibration from Kennicutt (1998). The relation found between the [OII] fluxes and the UV luminosity is in good agreement with the VVDS data and is valid for different galaxy populations. For the other emission lines, we adopt intrinsic, unextincted flux ratios of [OIII]/[OII] = 0.36; H beta/[OII] = 0.28; H alpha/[OII] = 1.77 and Ly alpha/[OII] = 2 (McCall et al. 1985; Moustakas & Kennicutt 2006; Mouhcine, et al. 2005; Kennicutt 1998).

The CMC reproduces the counts and color distributions of galaxies observed in bands from 0.4 to 2.3 microns, e.g., comparing to the GOODS (Giavalisco, et al. 2004) and UDF (Coe, et al. 2006) surveys. The CMC also provides an excellent match to the redshift-magnitude and redshift-color distributions for I<24 galaxies in the VVDS redshift survey (Le Fèvre, et al. 2005). For more detail about the CMC and its validation, see Jouvel, et al. (2009).

## 3.B Spectroscopic Sensitivity

To compute the sensitivities (exposure time required to obtain a certain S/N for a particular simulated galaxy spectrum, and redshift success rates), we start with input spectra from the DES galaxy catalog simulations (provided by R. Wechsler & M. Busha) or from the COSMOS Mock Catalog (CMC, Jouvel, et al. 2009), both of which model the detailed properties of the galaxy population at magnitudes and redshifts appropriate to DESpec. The DES catalog simulations provide galaxy spectra based upon the template models generated by the Kcorrect package (Blanton & Roweis 2007), while as just described the CMC provides spectra based on the COSMOS data, providing in particular detailed distributions of line fluxes for emission-line galaxies. We account for the transmission as a function of wavelength through the atmosphere.

For this initial study, we assumed that the DESpec system has 50% of the net throughput of DECam, accounting for losses in the fibers, grating, and other optics. As shown in Figure 3.1, this simple estimate accurately summarizes the total system throughput, based on early estimates for all system components, to within ~10% over the wavelength range of interest for most studies. Given the DESpec throughput, the simulations account for the fraction of galaxy light entering a fiber, using either an analytic Gaussian model and the galaxy size distribution from the CMC or the detailed DES galaxy image simulations. The dominant source of noise for the DESpec spectra comes from the night sky, which we model using the high-S/N, high-resolution spectral atlas of Hanuschik (2003) plus the extensive spectroscopic archives from the SDSS and BOSS surveys. We take care to use data with resolution similar to or better than DESpec, in order to properly sample the much lower sky continuum levels in between the forest of atmospheric airglow emission lines in the red.

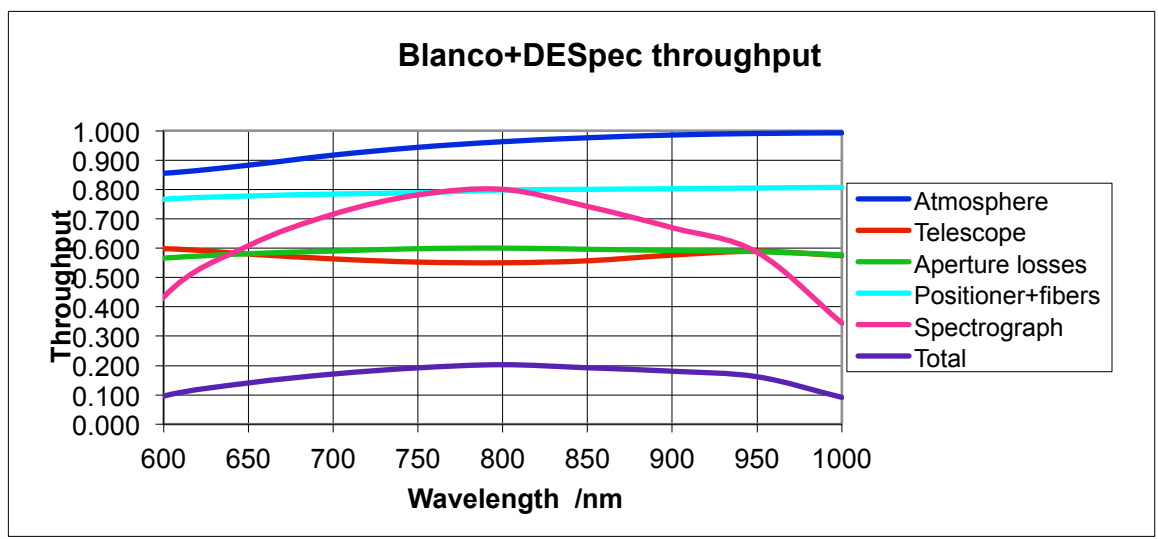

*Figure 3.1:* *Total Blanco+DESpec system throughput, conservatively assuming 50-meter fibers and single-arm spectrographs.*

The above components are implemented in a simulation package that generates flux-calibrated, one-dimensional extracted spectra with the signal and noise properties and wavelength resolution appropriate to DESpec. The simulated spectra are then used to calculate S/N and redshift success rates as functions of galaxy properties, in particular magnitude, redshift, and emission-line strength. For emission-line galaxies, we measure the S/N of the emission-line flux, especially of the [OII] 3727 line that is the most useful for redshift determination at $z$>0.5. We calculate the spectroscopic redshift success rate following Jouvel, et al. (2009): we require at least two lines to be detected at more than 5 sigma in the wavelength range of the spectrographs, or in the case of the OII line, we require that it is detected at greater than 5 sigma and that it is detected in its corresponding DES imaging band at 24th magnitude. In addition, we use the standard redshift measurement technique of cross-correlation with template spectra, as implemented in the IRAF external package `rvsao` (Kurtz & Mink 1998), to compare measured vs. true redshifts and to estimate the fraction of successful redshift measurements vs. galaxy properties. This will be useful for absorption-line galaxies, for which redshift measurement will be based upon spectral features spread over a range of wavelengths rather than on individual strong emission lines.

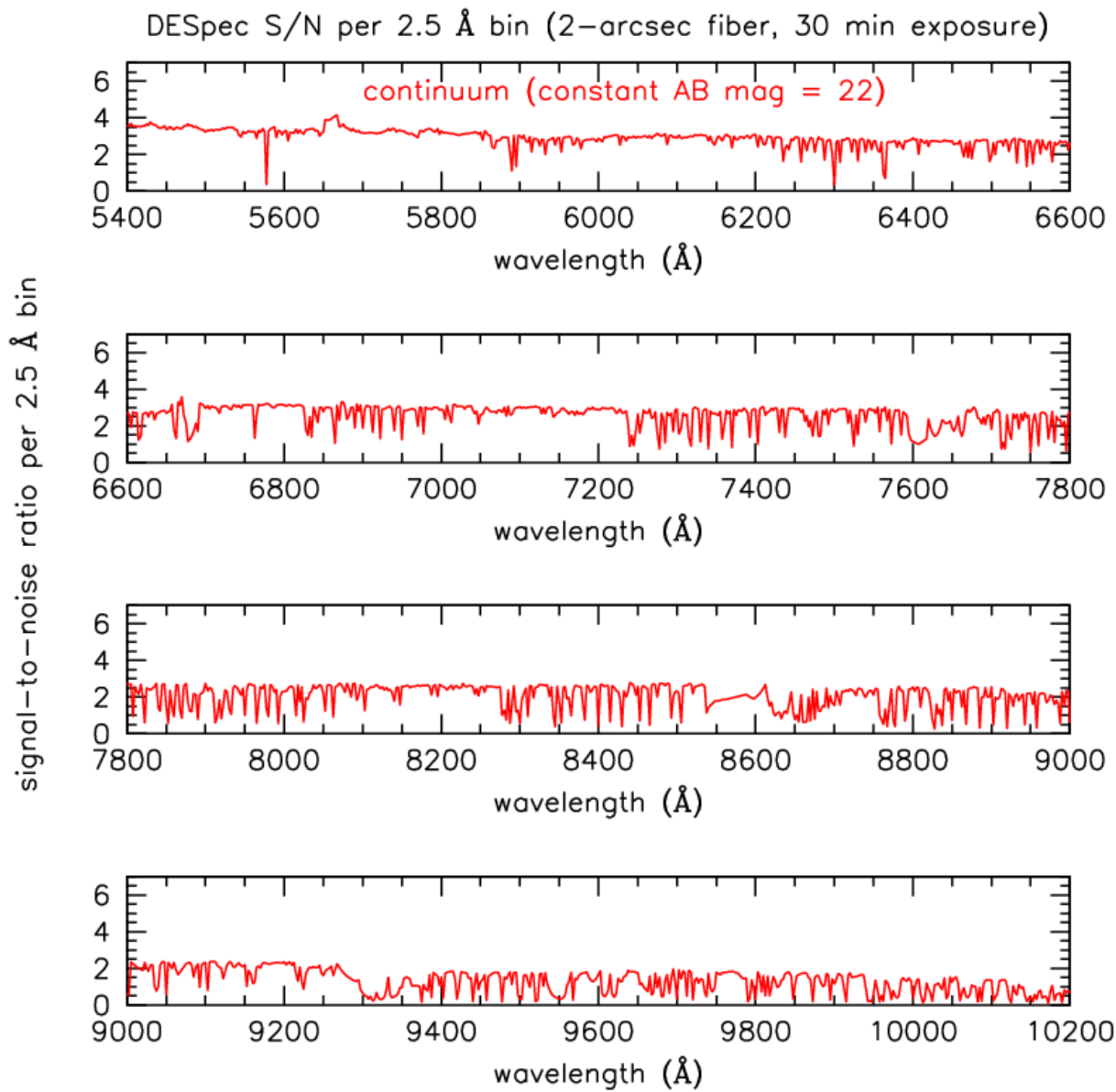

*Figure 3.2:* *The signal-to-noise ratio in 2.5 Å bins as a function of wavelength for an input spectrum with $m_{AB} = 22$ and an exposure of 30 minutes. The calculation assumes the DESpec throughput and sky noise as described in the text.*

We plot two examples of DESpec signal-to-noise calculations, one for an object with continuum level with constant AB mag = 22 (Fig. 3.2) and one for an emission line with flux = $1.0\times10^{-17}$ erg/s/cm$^2$ (Fig. 3.3). In both cases we assume a 2-arcsec-diameter fiber, 70% of the light entering the fiber, 30-minute exposure time, airmass 1.3, and the sky from Hanuschik (2003) as appropriate for dark time. We plot the S/N in fixed bins of 2.5 Å. In the following we adopt a flux limit of $3.0\times10^{-17}$ erg/s/cm$^2$ (three times larger than the flux illustrated in Fig. 3.3) and S/N = 5 for detection. For the emission line we assume a narrow line for which the flux is entirely contained within a single resolution element. As noted below in Section 4, the results are not very sensitive to the assumed fiber diameter.

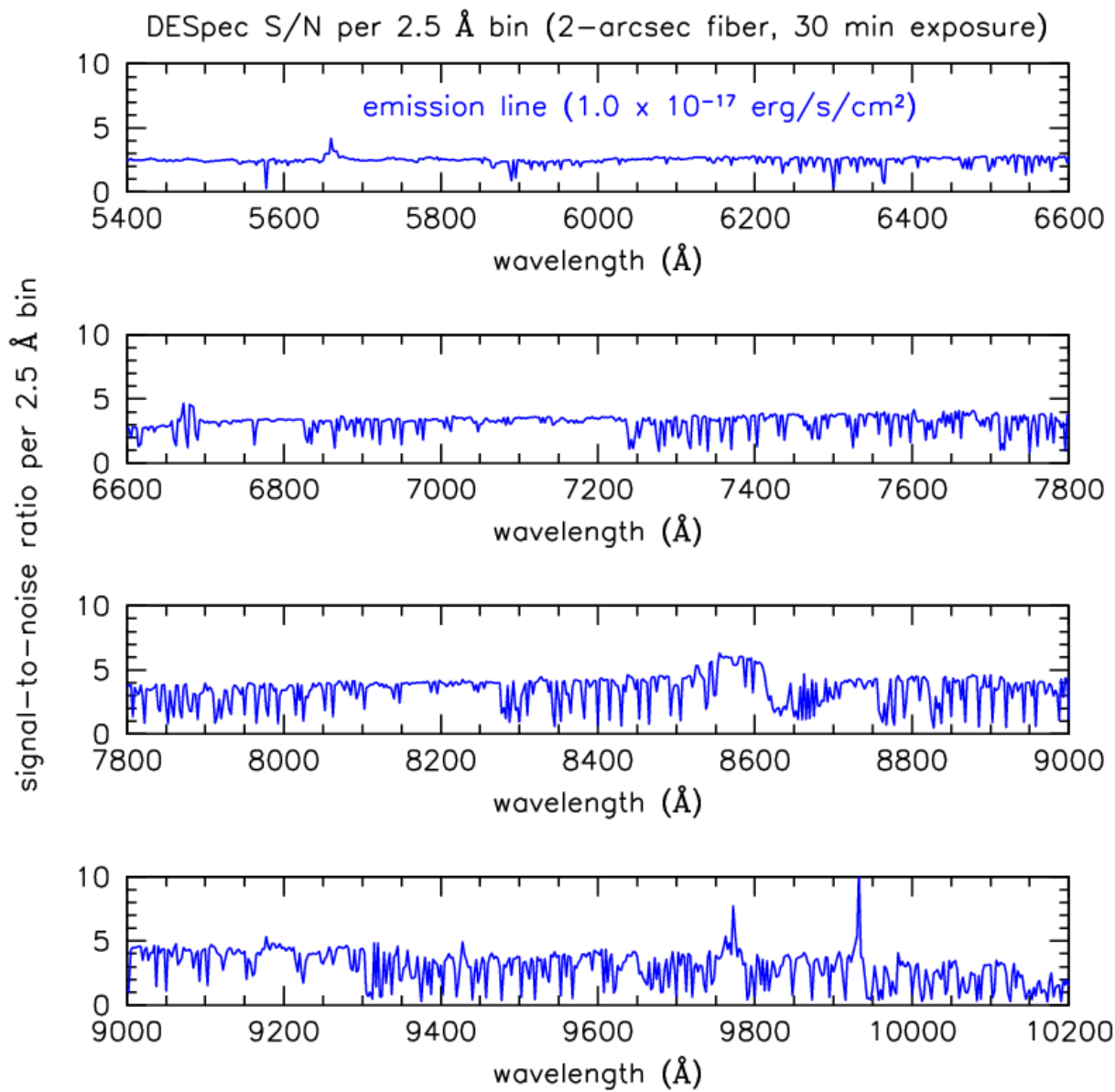

*Figure 3.3:* Same as Figure 3.2, except that the signal is an unresolved emission line with flux $10^{-17}$ erg/sec/cm$^2$.

## 3.C Luminous Red Galaxy Selection

For galaxy clustering measurements for baryon acoustic oscillations (BAO) and redshift-space distortions (RSD), LRGs are convenient because they occupy dense regions and are thus strongly biased relative to the dark matter, b~2, so the clustering amplitude is high. As noted above they are also important for cross-correlation with weak lensing of background galaxies. With our simulations we have estimated the completeness rate for objects in the CMC catalogue. We plot in Fig. 3.4 a preliminary example of the redshift completeness for LRGs, showing the magnitude, as a function of redshift, where the DESpec redshift success rate is expected to be 90%. (The 90% completeness magnitude gets fainter with increasing redshift, because strong absorption lines enter the DESpec spectral range with increasing $z$, which more than offsets the lower flux levels.) Also shown is the apparent magnitude vs. redshift relation for an LRG with luminosity of 3L* (Eisenstein et al. 2001), showing that the survey is expected to be about 90% efficient for bright (>3L*) LRG redshift measurement out to z = 1.3 in a 30 minute exposure, modulo target selection efficiency, fiber collisions, etc.

We have considered two different selection algorithms for LRG's. The first is based on optical and near-infrared data from the DES+VHS datasets. The second one is based on DES bands combined with VHS as well as WISE data. We have made two LRG selections for the following reasons: LRG's at z~0.5 usually inhabit massive dark-matter halos and hence are tracers of the dark matter field that produces the lensing measured by the DES sample. We therefore can extract extra cosmological information via cross-correlation techniques described earlier. The two selections are a proof-of-concept showing that these galaxies can be selected in large numbers at the appropriate redshifts both with a simple color cut and also much more efficiently with more complicated techniques (which might yield better results in terms of finding larger cross-correlations but might be harder to understand in terms of selection effects). We describe the selections below, including the number density of galaxies in each selection.

Using DES+VHS imaging, LRGs can be selected to yield a relatively flat redshift distribution over the range 0.5 < z < 1 by using selection cuts in *r-z* vs. *z-H* color space. We plot these color-color diagrams in Figure 3.5. These cuts supply more than the needed density of LRG targets, so the targets can be randomly sampled. The galaxies yielded by Cut I amount to 770 galaxies per square degree at magnitudes in the z band brighter than 21. For Cuts II and III, we have ~60 objects per square degree at magnitudes brighter than 22 in the z band, but they are at significantly higher redshifts. These cuts will also select some non-LRG objects.

The second selection criterion is based on a neural network method for selecting galaxies. Here we assume that we may select LRG's from the full multi-wavelength dataset including DES, VHS and WISE. In this selection, we classify galaxies as LRG's or not based on the spectrum from which they were created in the mock catalogue. We use a subset of around 10,000 galaxies to train the neural network to select galaxies that have spectra similar to those of LRGs based solely on the information provided by the colors. This selection method yields 1241 LRG targets per square degree at i< 22 with a photometric redshift larger than 0.5. We can see from Fig. 3.6 the distribution of selected galaxies vs. redshift from the mock catalog for these two selection methods. Also included in Fig. 3.6 for comparison is the number of galaxies that could be selected via PTF and WISE photometry (cf. BigBOSS). In all cases, the bulk of selected LRGs are in the desired redshift range 0<*z*<0.5 and the density of targets is higher than required.

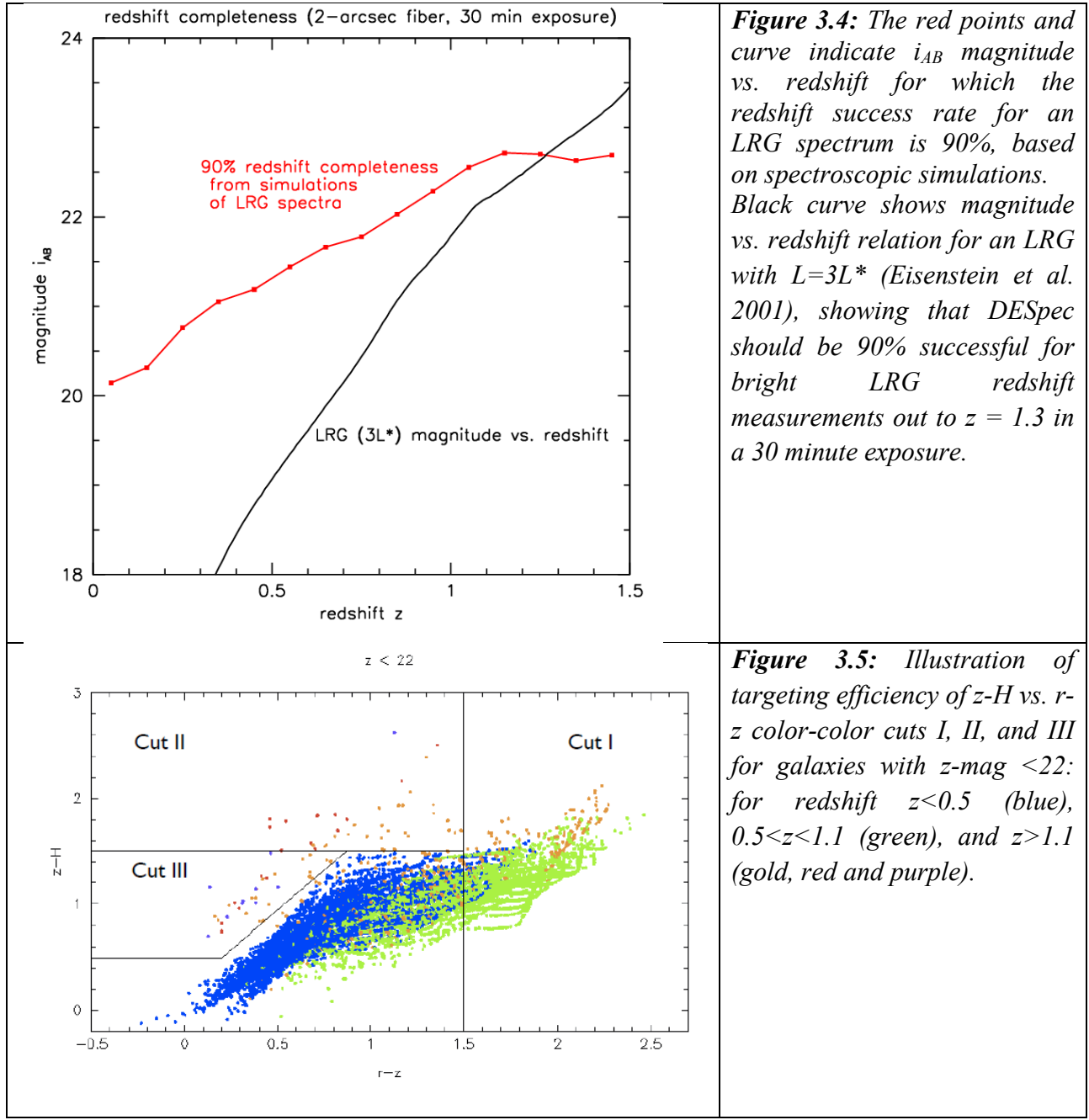

*Figure 3.4:* The red points and curve indicate $i_{AB}$ magnitude vs. redshift for which the redshift success rate for an LRG spectrum is 90%, based on spectroscopic simulations. Black curve shows magnitude vs. redshift relation for an LRG with $L=3L^*$ (Eisenstein et al. 2001), showing that DESpec should be 90% successful for bright LRG redshift measurements out to $z = 1.3$ in a 30 minute exposure.

*Figure 3.5:* Illustration of targeting efficiency of z-H vs. r-z color-color cuts I, II, and III for galaxies with z-mag <22: for redshift z<0.5 (blue), 0.5<z<1.1 (green), and z>1.1 (gold, red and purple).

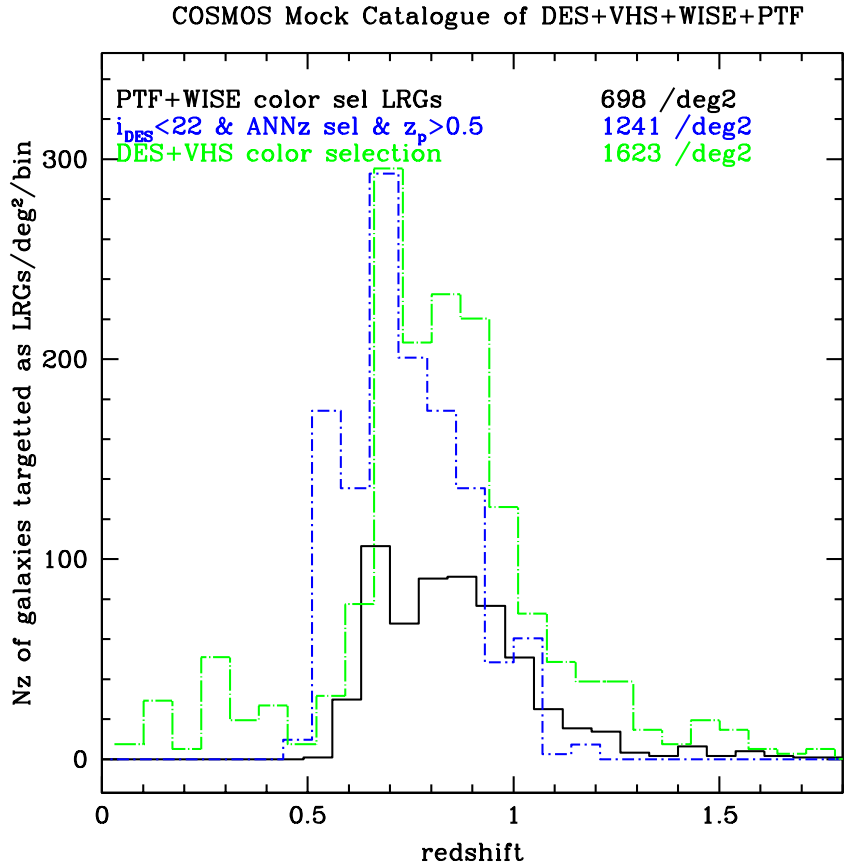

*Figure 3.6:* *Predicted redshift distribution for three methods of selecting luminous red galaxies based on the Cosmos Mock Catalog (see text).*

### 3.D Emission-line Galaxy Selection at High Redshift

Here we describe simulations to investigate selection criteria for a survey targeting emission-line galaxies at redshifts higher than accessible with luminous red galaxies, for BAO and RSD studies. Using the CMC (Jouvel, et al. 2009) to generate the flux and S/N for each emission line, the survey is expected to reach the line sensitivities described earlier, namely a signal-to-noise ratio of around 5 for fluxes larger than $3\times10^{-17}$ erg/s/cm$^2$ from wavelengths 600 to 1000 nm (Fig. 3.3). This is reached for a 30 min exposure, which we take as the baseline.

We can use photometric redshifts from DES + VHS to select high-redshift galaxies. Fig. 3.7 (left panel) shows the redshift distribution of emission-line galaxies that would be obtained with DESpec with a photometric redshift (using ANNz) cut between $1 < z_{phot} < 2$, i-magnitude<23, and a S/N=5 detection of one of various lines: OII line (blue), Hα (red), OIIIa (magenta), OIIIb (brown), or Hβ (gold). (Color-color instead of photo-z selection, similar to that of the DEEP2 redshift survey of ELGs over the redshift range 0.7 to 1.4, gives a similar result.) Exercises such as this give a first indication of the spectroscopic success rates, redshift distribution, and numbers of galaxies that would result from such a survey. A wavelength range 600–1000 nm allows Hα to be detected up to $z$=0.52 and [OII] to be detected at $z$>0.6; the upper limit in redshift for [OII] is $z$=1.7. We reach a total number of 2500 galaxies/deg$^2$ between redshift 1 and 1.7 for galaxies with i<23. Redshift measurements coming from lines other than [OII] provide an additional 300 gal/deg$^2$ in this redshift range. We can further increase the high redshift range of this selection by selecting slightly fainter galaxies and increasing the photometric redshift cut to higher redshifts. In Fig. 3.8 we have done such selection where some of the galaxies with photo-z closer to 1 have been removed from the selection. This selection has around 50 galaxies in a bin of 0.1 in redshift at redshift 1.7, which would be roughly enough to control the shot noise in a power spectrum measurement at these redshifts. In Fig. 3.8 we compare the redshift distribution of such a selection at i<23 and i<23.5 after further cuts in photo-z in order to increase the high-redshift tail with a simple color-cut selection based on three-optical-band data. This shows the extra redshift leverage obtained by deeper optical data in the selection process at the expense of having a more complicated target selection. These selections represent only a few percent of the galaxies available at those magnitude limits: we have ~ten galaxies per square arcminute available for selection, yet we are restricting selection to under one galaxy per square arcminute. Figure 3.8 also illustrates that selection with PTF + WISE photometry does not yield as many $z$>1 galaxies.

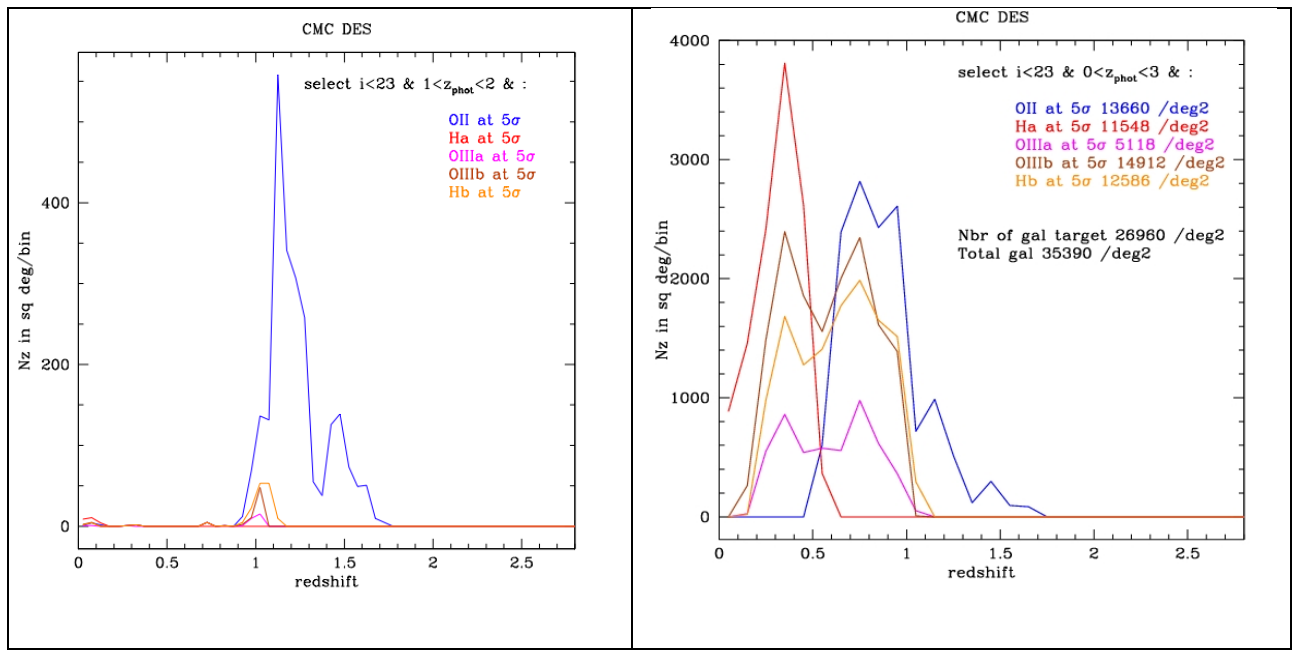

***Figure 3.7:*** *Example of the yield of a particular ELG target-selection scheme based on the COSMOS Mock Catalog (CMC) simulations, assuming 5-sigma line detections as expected for a 30-minute DESpec exposure. Galaxies brighter than i=23 are selected that have photometric redshifts between 1 and 2 (left); the distribution of those satisfying a much broader photometric redshift cut between 0 and 3 are shown in the right panel. This shows that selection of high-redshift sources via photometric redshift is effective and that, as expected, [OII] is the principal spectroscopic feature in this range of redshifts. Other redshifts are accessible with other lines, including Hα, Hβ, and OIII.*

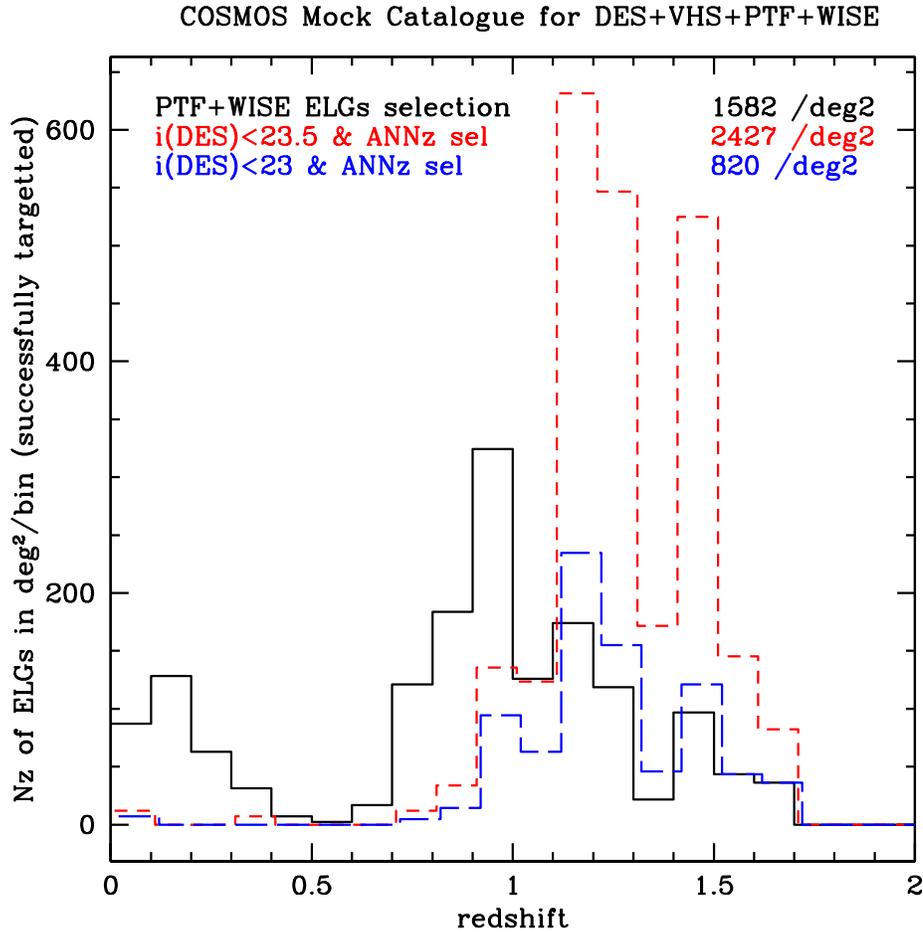

***Figure 3.8:*** *Predicted redshift distribution for three methods of selecting emission-line galaxies based on the Cosmos Mock Catalog.*

In Figure 3.9 we show photometric vs. true redshift for the same flux-limited and photo-z selected sample of galaxies. This allows us to estimate the galaxies we miss due to the photo-z selection. For galaxies selected with photometric redshifts between 1 and 2, 4.5% of the galaxies have true redshifts outside the desired redshift range of 1-2. If we choose instead to select a low-redshift emission-line sample (photo-z's between 0.5 and 1), 12% of the target galaxies have true redshifts outside the selection range. This study indicates that DES+VHS imaging can yield high-efficiency target selection of high-redshift galaxies based on photometric redshift cuts.

Neural networks were used in the above analysis to estimate the photometric redshift of a galaxy with a training set that includes the known redshift of the galaxy. We have also used neural networks to estimate which galaxies would be successful in yielding a redshift if targeted by DESpec. It has been shown that an efficient targeting strategy can indeed be realized using ANNz (Collister & Lahav 2004), and Abdalla, et al. (2008) showed that there is some predictability of emission line strength from galaxy colors. This is done in the following way: training-set galaxies with no emission lines were assigned the value zero and training-set galaxies with sufficiently strong emission lines were assigned the value one. Because of the Bayesian nature of ANNz, we can interpret the value generated by the trained network as a probability of each galaxy having yielded a successful redshift. (Strictly speaking, this is not a real probability, because it does not include the regularization function used to train the network.) Here we define the output of the network as the ANNz *target selection value* for that galaxy, ranging generally between zero and one. Using the CMC catalog in the DES+VHS bands, we show in Fig. 3.10 that we can select strong-emission-line galaxies via their target selection values.

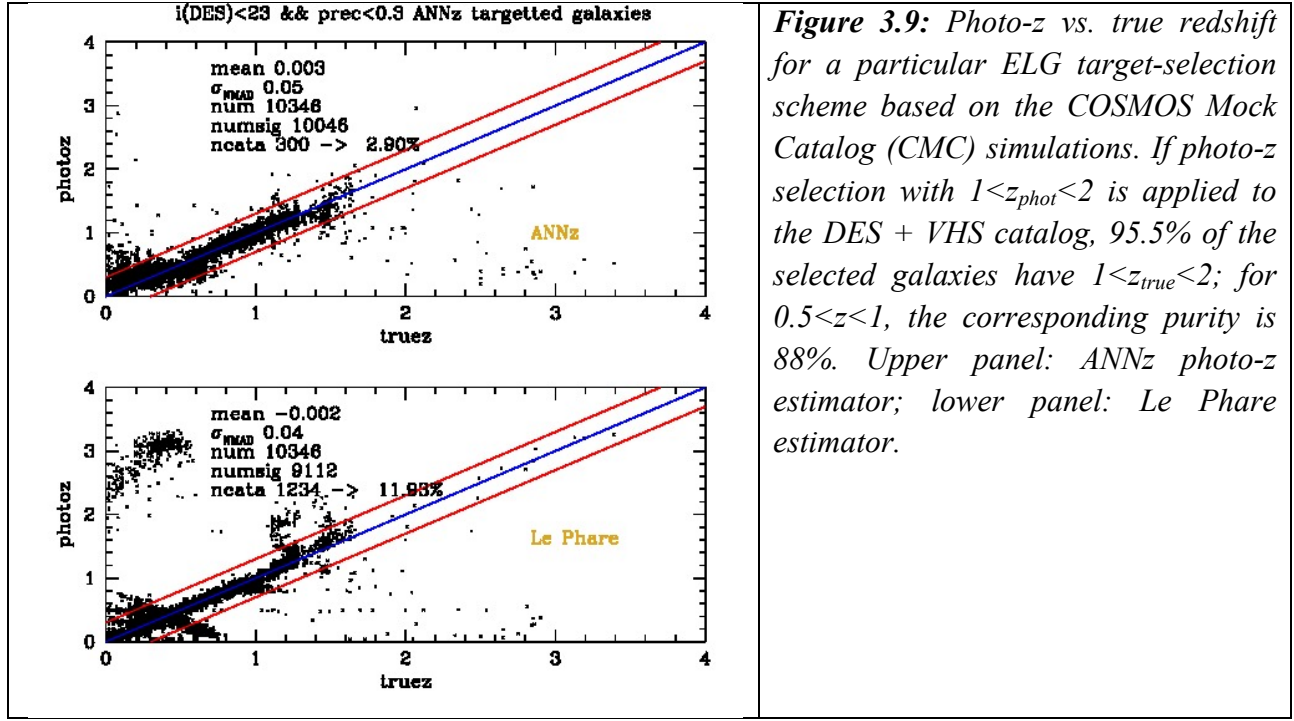

*Figure 3.9:* Photo-z vs. true redshift for a particular ELG target-selection scheme based on the COSMOS Mock Catalog (CMC) simulations. If photo-z selection with $1<z_{phot}<2$ is applied to the DES + VHS catalog, 95.5% of the selected galaxies have $1<z_{true}<2$; for $0.5<z<1$, the corresponding purity is 88%. Upper panel: ANNz photo-z estimator; lower panel: Le Phare estimator.

Using 10,000 galaxies as a training sample, we derive ANNz target selection values. In a real survey we would have to spend a few nights obtaining a training set by targetting galaxies in a wider color range to make sure that our selection is not slightly biased from possible miscalibrations of our mock catalog. Obtaining this training set with DESpec would be feasible in only a few nights. To define a selection criterion, we plot the cumulative number of galaxies as a function of the target selection value as shown in Fig. 3.11. A cut at a value of 0.8 selects most of the emission-line galaxies and has small contamination from weak-emission-line galaxies. We then derive the Spectroscopic Success Rate (SSR) as the percentage of galaxies for which we will be able to measure a redshift (as before, based on the significance of detection of an emission line of given flux), as shown in Figs 3.11 and 3.12. The dotted lines in Fig. 3.12 represent the SSR for different magnitude cuts, while the solid lines represent the percentage of galaxies compared to the total number of galaxies at the magnitude cut.

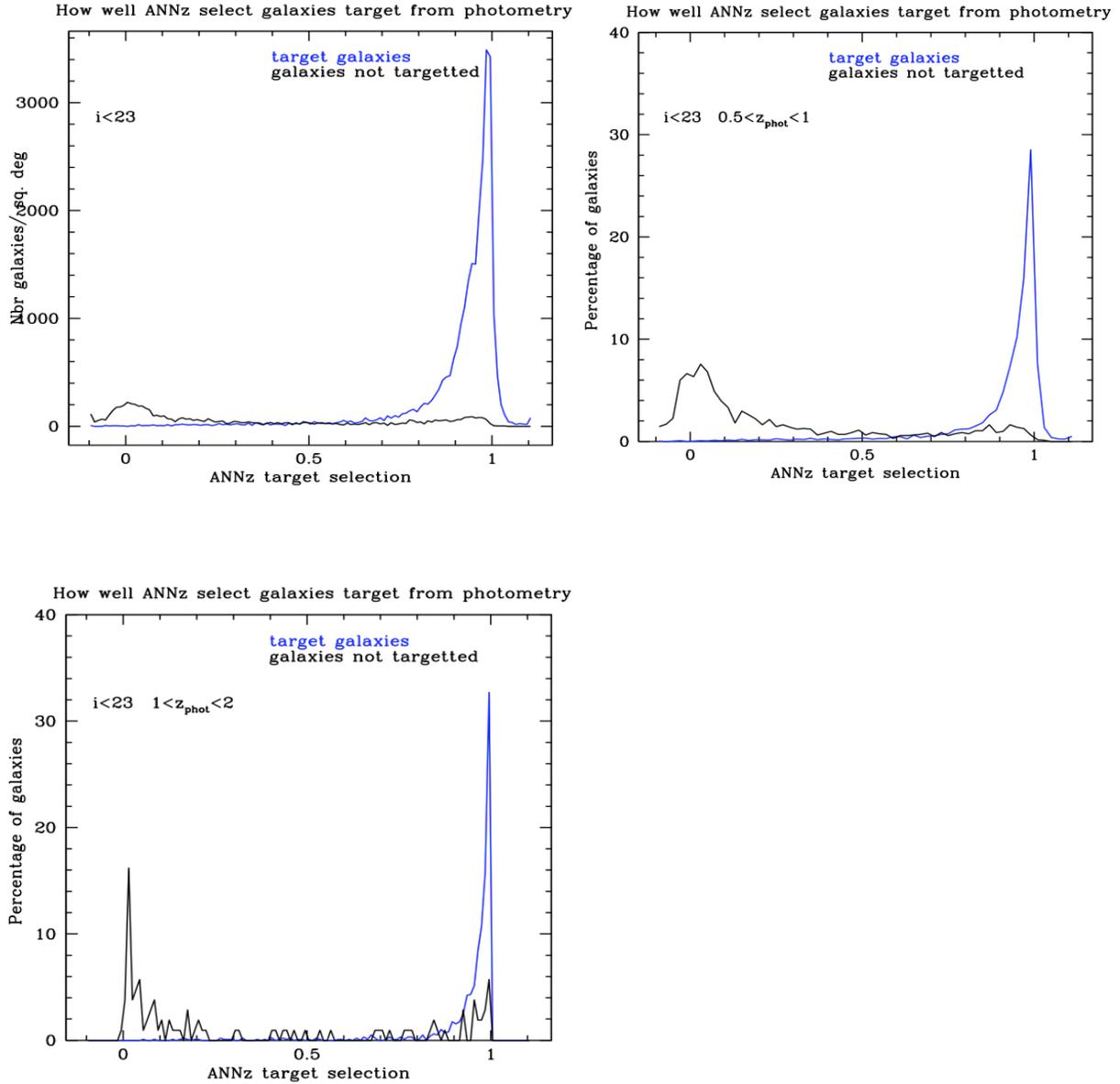

*Figure 3.10:* *Top left panel: number of galaxies per square degree as a function of the ANNz target selection value, for galaxies with i<23. Blue line corresponds to galaxies for which we can measure a redshift, the "target galaxies:" they have at least one emission line detected at 5 sigma using DESpec, and we assign them a value of 1. Black line corresponds to galaxies that do not have a strong emission line detectable; we assign them a value of 0. Top right panel: same, but with additional photo-z selection cut of 0.5<$z_{phot}$<1. Bottom panel: same, but with photo-z cut of 1<$z_{phot}$<2.*

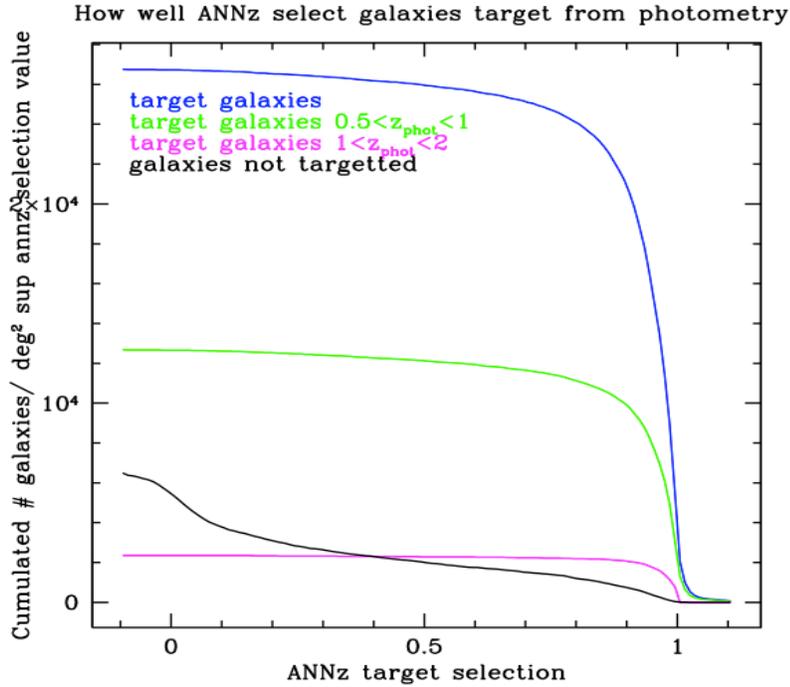

*Figure 3.11:* *Cumulative number of galaxies/deg$^2$ as a function of the ANNz target selection value for target galaxies in blue, target galaxies with a photo-z between [0.5-1] in green, and [1-2] in magenta, and for galaxies not targeted in black. Using an ANNz target criterion of 0.8, we are able to reach a SSR of 95% at i<23 AB mag, targeting close to 75% of all galaxies at i<23. This yields 24,000 galaxies/deg$^2$.*

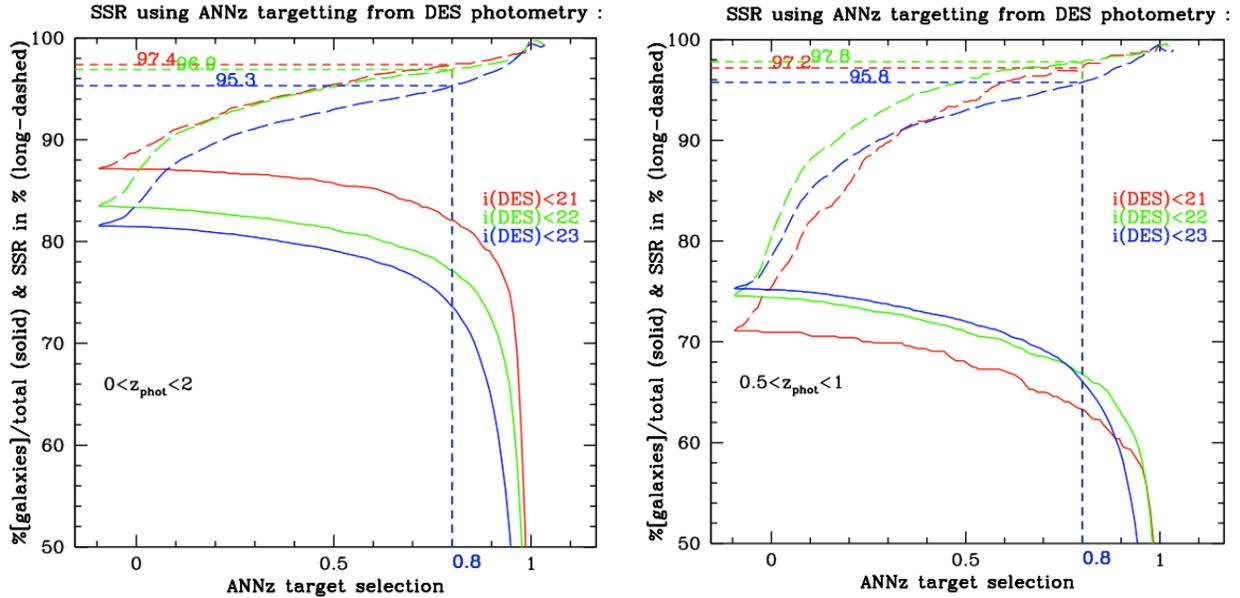

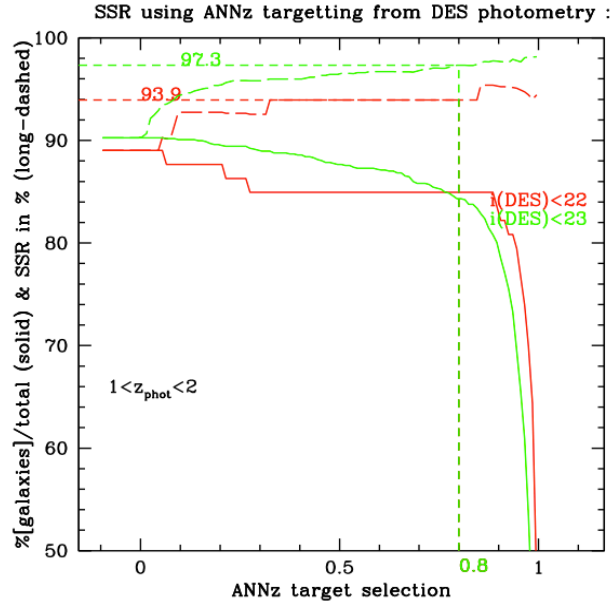

*Figure 3.12:* Spectroscopic Success Rate (dashed lines) and cumulative number of galaxies (solid lines) as a function of ANNz target selection values for magnitude i<23 in blue, i<22 in green, and i<21 in red. Top left figure represents all galaxies meeting the different magnitude cuts, top right figure is for galaxies with 0.5 <$z_{phot}$< 1, and the bottom figure for 1< $z_{phot}$ < 2.

### 3.E Survey Strategy

This section gives an outline of how the survey might acquire the spectra. The constraints are the field of view of the focal plane (3.8 sq. deg), the number of fibers (~4000), the fraction of fibers reserved for community use, and the total survey time available. Assuming the fibers fill the circular field-of-view, the fiber density on the sky is 1050 targets per sq. deg, about 70% of the desired surface density of 1500 successful galaxy targets per sq. deg. for BAO and RSD studies to redshift z ~ 1.5 (Secs. 2.B and 2.C). Therefore, including realistic rates for the fraction of fibers allocated to targets and the redshift completeness, we need to cover the sky about twice in order to achieve the desired surface density. To cover each part of the 5000 sq. deg. DES area at least twice requires 3200 pointings. Integrations of ~30 minutes per field are required, or about 40 minutes total time per field including overheads. Assuming 80% of the allocated nights are astronomically useable, and an average of 7.7 hours on the sky per useable night (both based on historical CTIO weather data) leads to a requirement for about 350 nights.

Assuming the sky is hexagonally tiled and all fibers within each hexagonal tile are given to our survey, then the number of ELG/LRG targets per field is 3300, with a mix of ~ 2700 ELG, 400 LRG, and 200 sky fibers, leaving 700 fibers for community programs. Assuming a redshift completeness of 75% for ELGs and 90% for LRGs gives a final sample of 7.6 x $10^6$ galaxies (6.4M ELGs, 1.2M LRG's). These numbers scale directly with the fraction of fibers given to the survey, with the balance being available for community programs.

Since there are of order 10,000 galaxies per square degree brighter than i = 22.5, tile-to-tile fluctuations due to large-scale structure can be accommodated by small variations in the selection criteria (e.g. the limiting magnitude, which can be readily incorporated into any analysis) or by the tile overlaps.

A further 700 fibers/field remain available for additional targets or repeat observations, in the tile overlap regions, which would allow 2 million additional spectra to be taken. But since these overlaps are observed 4 times, much longer integration times can be built up for these additional targets, allowing fewer but fainter sources to be targetted. The AAO is investigating a smaller positioner pitch, ~6 mm (Sec. 4.B). If this is implemented, and the field-of-view restricted to a hexagon but still with 4000 fibers, then the above numbers are maintained, but the additional sources can then be uniformly distributed over the sky.

Looking forward to the LSST era, one could cover 15,000 sq. deg. (the bulk of the LSST extragalactic sky) in ~1000 allocated nights, obtaining ~23 million ELG/LRG redshifts and up to ~6 million additional spectra.

The impact of altering the observed sample from this strawman plan is demonstrated by some initial results in Section 3.G. At present we have limited ourselves to two populations, Luminous Red Galaxies (LRGs) down to a magnitude of i<22 and Emission Line Galaxies (ELGs) down to a magnitude of i<23.5, and only alter the ratio of LRGs to ELGs. We consider three cases of LRG/ELG: 10%/90%; 33%/67% and 50%/50%. We assume that the whole survey area adopts this ratio for the entire observing time. A number of more detailed results, which allow the type of observed galaxies to vary with redshift and observed position on the sky, will be presented in Jouvel, et al. (2012).

## 3.F Projected Angular Power Spectra Analysis

In this section we introduce forecasts for joint constraints from DES+DESpec using projected angular power spectra, *C(l)*, in redshift shells, applying the above selection criteria. This technique is common in weak lensing and photo-z work, but spectroscopic galaxy surveys usually use a full 3D power spectrum analysis. While the *C(l)* approach can lose small-scale information through projection effects, the benefit is a very elegant way of including cross-correlations between weak lensing and galaxy clustering, including all the off-diagonal elements of the covariance matrix. On scales of interest for our forecast, Asorey, et al. (2012) have shown that this approach can in fact recover the full 3D clustering information. It is also much simpler to examine the impact of the galaxy distribution function, *n(z)*, on galaxy clustering measurements, which is important when considering target-selection strategies. This approach is complementary to the work of Cai & Bernstein (2012) and Gaztanaga, et al. (2012) described in Section 2.

We assume two fiducial survey designs. We model DES as a cosmic shear survey with photometric-quality redshifts, measuring cosmic shear of ~300 million galaxies over 5000 deg$^2$ out to redshifts beyond 1.5. We assume DESpec will measure spectroscopic redshifts for ~10 million galaxies over 5000 deg$^2$ out to redshift 1.7. We analyze the DES survey using five unequally spaced tomographic bins in redshift with roughly equal galaxy number density in each. The higher quality redshifts in the DESpec survey allow us to use 30 bins of equal redshift spacing. We include the effects of Redshift Space Distortions in our *C(l)* framework using the formalism introduced by Fisher et al. (1994). This study includes all of the *P(k)* information, binned in *l* and *z*, hence includes some BAO information. We examine each survey independently, forecasting constraints through the Fisher Matrix (FM) formalism, as well as joint constraints from both surveys assuming that they observe different patches of sky (simply adding the FMs) or the same patch of sky (including all galaxy-shear cross-correlations). We exclude nonlinear scales from all observables that include galaxy clustering. Galaxy bias, $b_g(z)$, is assumed to be scale-independent with fiducial value 1, and is parameterized by an overall amplitude normalization and three nodes in redshift (similar to Gaztanaga, et al. 2012). The amplitude of these four nuisance parameters is allowed to vary and is marginalized over in our results. For galaxy-shear results, the cross-correlation coefficient is assumed to be unity at all scales/redshifts, $r_g(k, z) = 1$, as indicated by simulations (Gaztanaga, et al. 2012).

### 3.G Results for Figure of Merit vs. Survey Strategy

Here we present the results of our projected angular power spectra analysis of DES+DESpec for both our fiducial survey strategies and for the simple alternative target selection scenarios mentioned above in Sec. 3.E.

Results are presented for both the standard Dark Energy (DE) parameters as well as two common Modified Gravity (MG) parameters, $Q_0$ and $R_0$, that parameterize deviations from General Relativity (GR). $Q_0$ parameterizes deviations from the standard Poisson equation, while $R_0$ is the ratio between the two metric potentials, $\phi$ and $\psi$. Both are unity in GR; see Bean & Tangmatitham (2010) for details of this parameterization. This two-parameter model has the advantage of being able to detect the presence of anisotropic stresses on cosmological scales but is otherwise complementary to the one-parameter MG model discussed above in Sec. 2. In all cases, a suite of standard cosmological parameters are marginalized over as well as nuisance parameters allowing for uncertainty in galaxy bias as a function of redshift. Results including galaxy clustering are restricted to linear scales.

Fig. 3.13 shows 95% likelihood contours from our Figure of Merit forecasts for both DE and MG parameters for DES cosmic shear (γγ), DESpec galaxy clustering (*nn*), DES+DESpec γγ + *nn* (different sky), and DESxDESpec γγ + *n*γ + *nn* (same sky, including galaxy-shear cross-correlations). These results are based on the naïve assumption of a flat redshift distribution out to z = 1.7; in Table 3.1 below we assume the more realistic redshift distribution given by the simulated target selection described in Sections 3.C and 3.D. Here we do not include priors from Planck or Stage II DETF experiments.

It is clear that combining both probes (weak lensing and galaxy clustering) improves the constraining power on Dark Energy and on Modified Gravity. Taking the example of LRG/ELG = 33/67 in Table 3.1, going from DES γγ to DES+DESpec γγ + *nn* improves the DETF FoM by a factor of 3.1. What is also very evident is that the additional cross-correlation information gained from having two surveys on the same patch of sky is very significant. The improvement in going from DES+DESpec γγ + *nn* to DESxDESpec γγ + *nγ* + *nn*, i.e., from different skies to same sky, is a factor of 1.5 in the DETF FoM. This is understandable when we consider the extra control of galaxy biasing provided by the galaxy-shear cross-correlations.

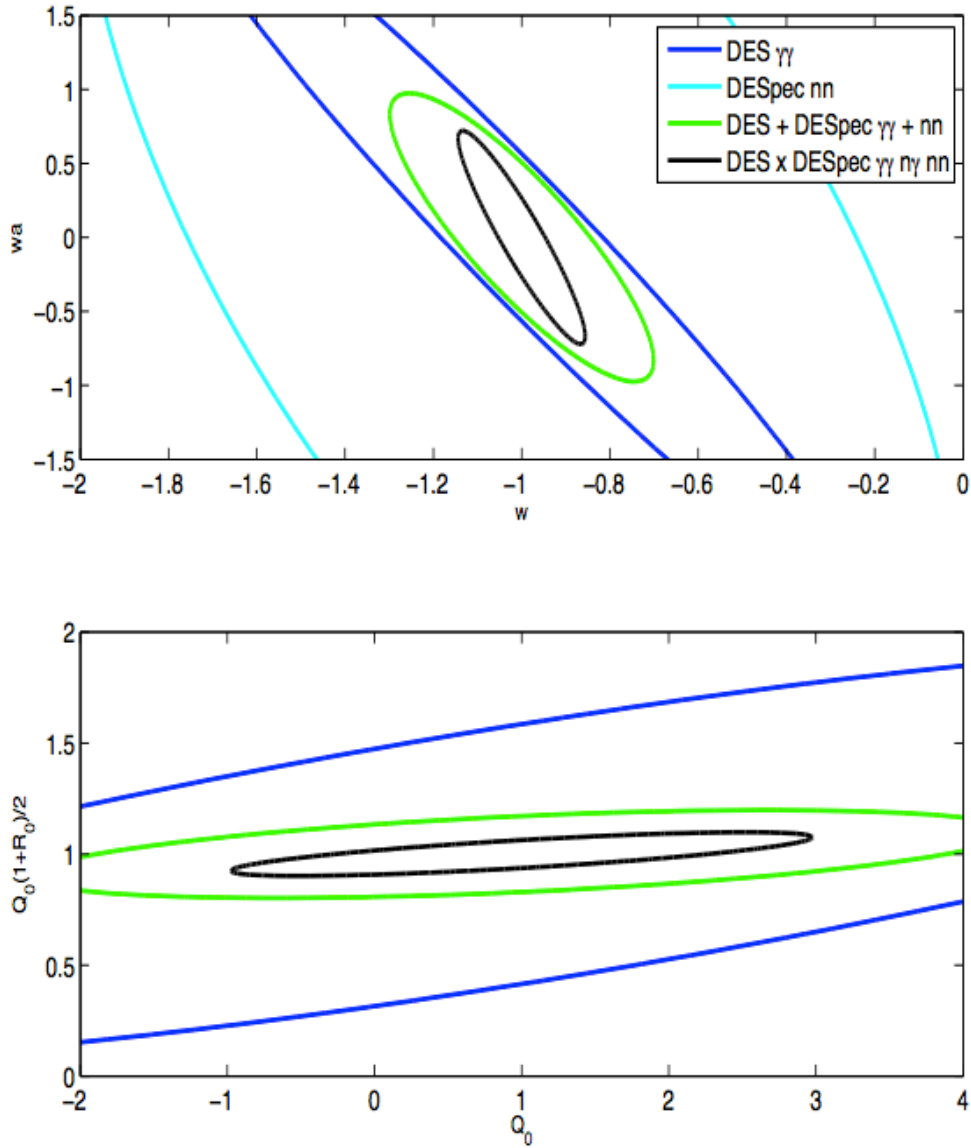

*Figure 3.13:* 95% confidence contours for DES cosmic shear (γγ) [blue lines], DESpec galaxy clustering (nn) [cyan lines], DES+DESpec γγ + nn (different sky) [green lines], and DESxDESpec γγ + nγ + nn (same sky) [black lines]. Top panel: Constraints on the equation of state of dark energy, $w_0$ and $w_a$, assuming standard GR. Bottom panel: Constraints on deviations from GR parameterized by $Q_0$ and $R_0$; all other cosmological parameters (e.g., $w_0$ and $w_a$) and nuisance parameters are marginalized over. In both cases, no Stage II or Planck priors are assumed. These results are initial findings and a full investigation will appear in Kirk, et al. (2012).

| Target Selection Scenarios FoMs | | | |
|---|---|---|---|
| LRG/ELG ratio | $nn$ | $nn + \gamma\gamma$ | $nn + n\gamma + \gamma\gamma$ |
| DE 10/90 | 1.00 | 3.51 | 5.22 |
| DE 33/67 | 1.30 | 4.05 | 6.09 |
| DE 50/50 | 1.17 | 3.82 | 5.73 |
| MG 10/90 | 1.00 | 3.47 | 3.86 |
| MG 33/67 | 1.29 | 4.01 | 4.65 |
| MG 50/50 | 1.68 | 3.78 | 4.33 |

*Table 3.1. Relative Figures of Merit for both DE and MG as a function of LRG/ELG ratio. DE and MG constraints are each independently normalised to the FoM for galaxy-galaxy clustering (including RSD) from a DESpec-type survey only. Columns show constraints for DESpec alone (nn), joint independent DESpec and DES constraints (nn + γγ), and joint constraints assuming DES and DESpec are in the same patch of sky, including all cross correlations (γγ + nγ + nn). Rows show results for different ratios of LRGs to ELGs as target selection strategies are changed. No Stage II or Planck priors are included here.*

The results for MG are similar. If we define an analog to the DETF FoM for MG (roughly the inverse of the area under the $Q_0$, $Q_0(1+R_0)/2$ contour), again for the instance LRG/ELG = 33/67, we see a factor of 3.1 improvement in going from DES γγ to DES+DESpec γγ + nn. The strong degeneracy between the metric potentials, ϕ, ψ, can be broken by joint use of weak lensing and galaxy clustering. Even so, there is still some improvement in going from different sky to same sky.

Table 3.1 summarizes the FoM variation as a function of LRG/ELG ratio. Even for the simple scenarios we consider, altering the distribution of galaxy types selected can change FoM results by up to 70% for DESpec alone. The addition of the (unchanging) DES survey mitigates this variation in the joint constraints, but we still see differences at the 20% level for the joint constraints including all cross-correlations. These results are preliminary findings, but they demonstrate the importance of considering the details of target selection for a DESpec-type survey. A more detailed investigation will appear in Kirk, et al. (2012).

In summary, Fig. 3.13 and Figs. 2.1 and 2.3 illustrate the potential for substantial improvement in our knowledge of DE and MG with respect to what will be achieved by DES. The combination of DES and DESpec can improve the DETF figure of merit on cosmic acceleration by large factors, depending on the probes and assumptions used (e.g., a factor of 6.09/1.30 = 4.7 in the case of DE, LRG/ELG = 33/67 from Table 3.1). A similar gain can be achieved simultaneously on the cosmic growth history or modified gravity models. We have also illustrated how we can use DES photometric information to select spectroscopic targets to optimize the survey science.

These results have assumed basic target selection strategies for a DESpec-type survey. We are pursuing R&D work on survey optimization. In particular we will optimize the relative number of galaxy targets of a given type at different redshifts (see Table 3.1). Different galaxy types differently trace dark matter halos of different mass. Higher-redshift galaxies sample larger volumes, but 3D dark matter information is better recovered at intermediate redshifts, where DES weak lensing information peaks (see Fig. 2.2). We will combine these theoretical considerations with those related to the spectrograph design to forecast results based on different design specifications, particularly wavelength coverage. This more detailed work is in progress and will be presented as part of the continued R&D development of DESpec.

## 4. DESpec Instrument Definition

DESpec will be a multi-fiber spectrometer that uses the same mechanical structure and most of the same corrector optics as DECam, thereby achieving a substantial cost savings compared to an ab-initio instrument. The instrument will be able to observe ~4000 targets simultaneously over a 3.8 sq. deg. field of view. As we are still exploring the science and survey requirements at this time, we are considering two designs. The first is a single-arm spectrograph with a wavelength range $550<\lambda<950$ nm. The second is a two-arm (dichroic) spectrograph in which the blue side has wavelength range $480<\lambda<780$ nm and the red side covers $750<\lambda<1050$ nm. For the single arm or the red arm of the two-arm design, the spare DECam CCDs could be used. We are also investigating the value of including an Atmospheric Dispersion Corrector (ADC). DESpec would be interchangeable with DECam in an acceptably short time. For our reference design we are adopting the two-arm design without an ADC. More detailed descriptions of each design are presented below.

## 4.A DESpec Optical Corrector Design

The DECam optical corrector was designed to make excellent images in individual filter bandpasses, but refocus was allowed, and lateral chromatic aberrations were controlled only within each bandpass. The 5-element DECam corrector optics produces a flat focal plane with a clear aperture radius of 225.54 mm. It has an f/2.9 beam that has 3.8 degrees of non-telecentricity at the edge of the focal plane. It also uses a filter-changer with 13 mm-thick filters. The DECam lenses are called "C1" (the largest, furthest from the focal plane) through "C5". C5 is the final DECam optical element and is also the imager Dewar window, so it travels with DECam when it is removed from the telescope. C1 through C4 would remain in the prime focus cage during an instrument swap of DECam with DESpec. As of this writing (August 2012), the assembled and aligned DECam corrector optics have been installed and aligned on the Blanco telescope.

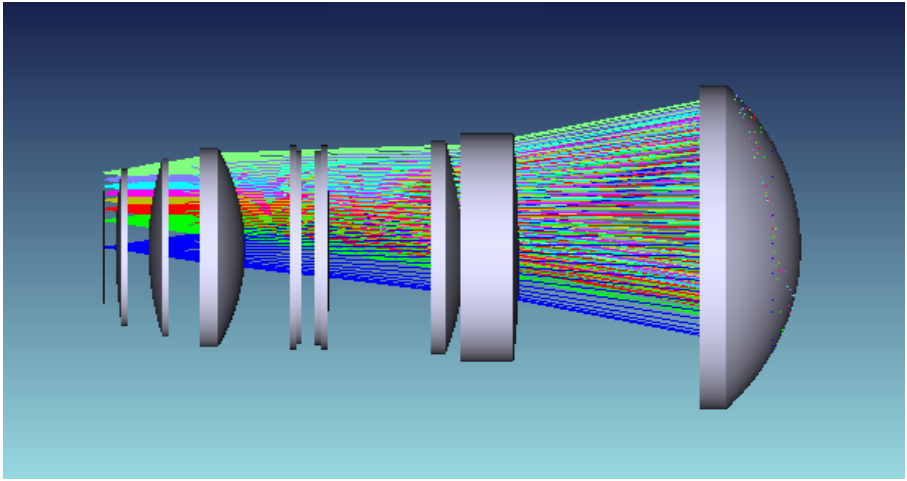

***Figure 4.1:*** *One option (DESpec-SK-3C) for the DESpec optics. From right to left they are C1 to C3, the two-component ADC, C4, C5', and the field-flattener C6. The focal plane of fiber-ends would be just to the left of the new C6. "C1" is a little less than 1 meter in diameter. The optical train is 1.9 meters long. C1, C2, C3, and C4 are already in place for DECam.*

4.A.1 DESpec Optical Corrector

DECam was required to deliver images that were comparable to the median site seeing. For DESpec, images are required to fall within a fiber diameter that is optimized for maximal S/N ratio of faint galaxies along with a minimum required spectral resolution. While the requirements on image sizes are similar, the impact of degradation in image size is different in the two cases. For DECam, a degradation of image size results in an inability to detect the smallest and faintest galaxies. For DESpec, a degradation of image size results in a reduced S/N ratio, but one can compensate (within limits) by increased observing time. DECam placed no constraint on the telecentricity of the beam incident on the focal plane, and the incoming beam is tilted up to 3.8 degrees at the focal plane edge. For DESpec, the fibers are constrained to be perpendicular to the focal plane, and an inclined beam would cause focal ratio degradation at the exit of the fiber. Thus, the beam needs to be perpendicular to the focal plane (telecentric) at all locations.

In addition, the DESpec corrector is required to produce a good image (though not as good as DECam) over the entire useful wavelength range. DESpec will reuse the 1$^{st}$ four elements of the DECam optical corrector (C1-C4). As noted above, C5 as well as the DECam filters would not be used for DESpec. For observations at high zenith angles, an Atmospheric Dispersion Compensator (ADC) can compensate for the natural prismatic effect of chromatic refraction in the atmosphere. Because the maximum zenith angle of DESpec has not been finalized, we have developed options for corrector optics both with and without an ADC. Here we present the option with the ADC, and in the next subsection we describe the ADC.

The single C5 lens in DECam is replaced with a pair of lenses C5' and C6 in DESpec. Both are made of fused silica. Such a pair is needed in order to achieve proper focus and telecentricity simultaneously. One surface (the concave face of the new C5') is aspheric. By using an asphere, the image quality is significantly improved at the field edge, and the curvature of C5 can be significantly reduced. Both lenses are rather thin, and the presence of an asphere on one might be of some concern. The thinness is somewhat to compensate for the extra glass thickness introduced by the ADC. However, the spectroscopic corrector lens of the SDSS 2.5 m telescope is even thinner and has a more severe aspheric shape on its convex side, so fabrication is expected to be feasible.

The present default DESpec corrector design, "DESpec-SK-3C", is shown in Figure 4.1. The optics achieves good spot size for wavelengths 500 nm < $\lambda$ < 1050 nm, as shown in Figure 4.2. The RMS spot radius is 0.25" at the center and 0.44" at the edge. The peak off-incidence ray (non-telecentricity) is at a 0.45 degree angle of incidence. The focal surface has a radius of curvature of 8047.2 mm. The focal ratio of the corrector, f/2.9, is in the optimal range for collecting light in an optical fiber (Ramsey 1988).

A DESpec corrector design without an ADC has better panchromatic image quality at Zenith, but suffers from chromatic refraction at other pointings, with the crossover being at ~40 degrees.

4.A.2 ADC Design

The atmospheric dispersion compensator is composed of a pair of crossed Amici prisms. Each prism is itself a double prism made of a crown and a flint glass with similar refractive indices. We have chosen N-BK7 and LLF1, since these glass types are typically selected for use in other large ADCs. By selecting the rotation angle of the crossed prisms, one can compensate for the dispersive effects of the atmosphere up to zenith angles of 60 degrees. The ADC designed for the WIYN One-Degree Imager (Muller, 2008) is similar in size and design to that which we expect for DESpec (see Figure 4.3). Optical studies show that improved images at the edge of the field can be obtained by making the first and last surfaces of the ADC mildly curved. Such an approach has been used for previous ADCs (e.g., the current Blanco prime focus corrector) and recent conversations with vendors indicate that they are not expected to be difficult to manufacture.

DESpec's ADC and shutter will fit into the large slot in the barrel that DECam uses for the filter-changer and the same shutter. There is sufficient room for an ADC in the DECam Barrel at the position of the filter-changer/shutter assembly, which is removable or installable in a short time. We do not need to use the filter changer while doing spectroscopy.

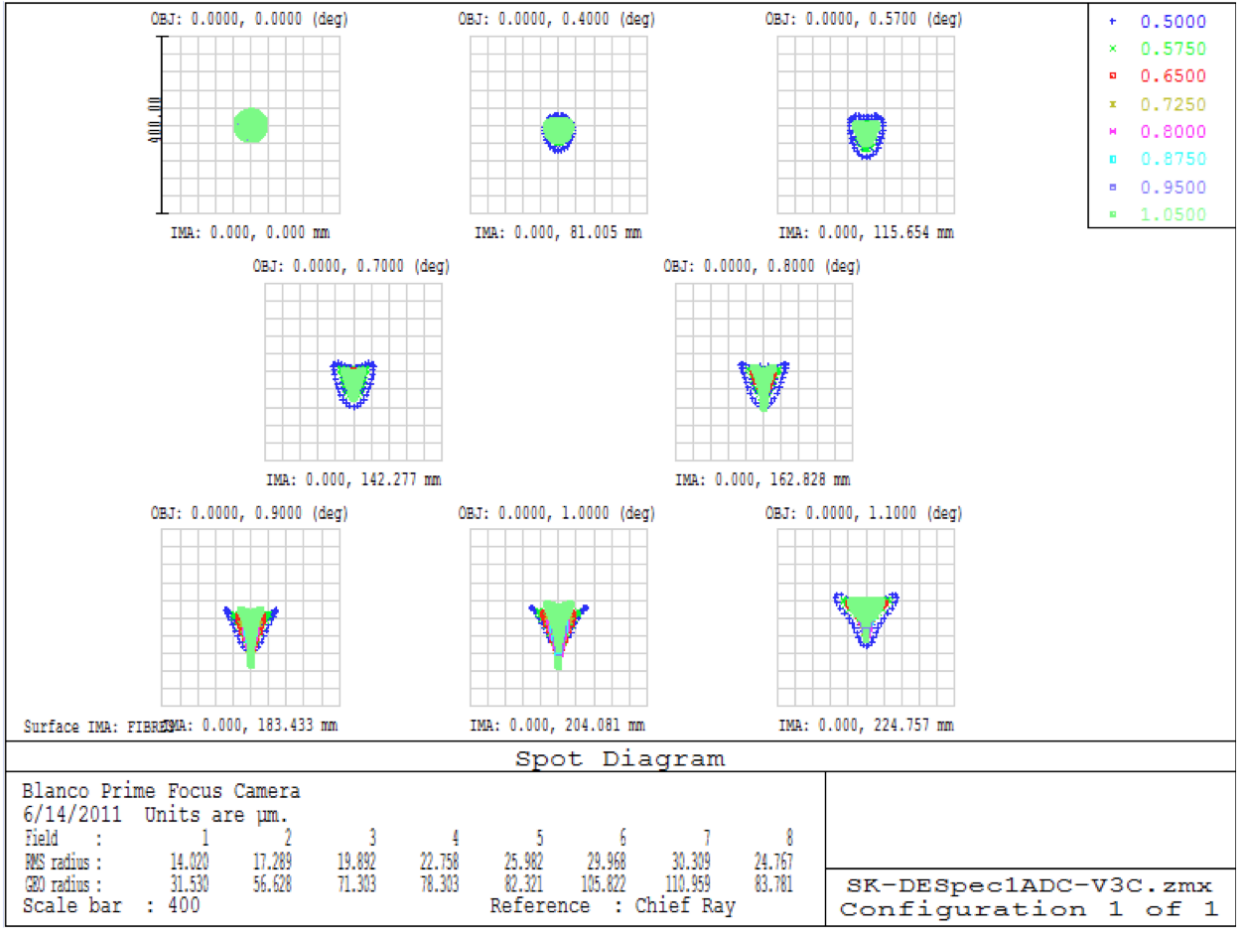

***Figure 4.2:*** *The spot size versus wavelength (0.5 (blue)–1.05 (green) microns) from the center of the focal surface (top-left) to the edge of the focal surface (radius of 1.1 deg, lower-right). The RMS spot radius is 0.26" at the center, 0.52" at the worst radius, and 0.44" at the edge. These results are from the "DESpec-SK-3C" design.*

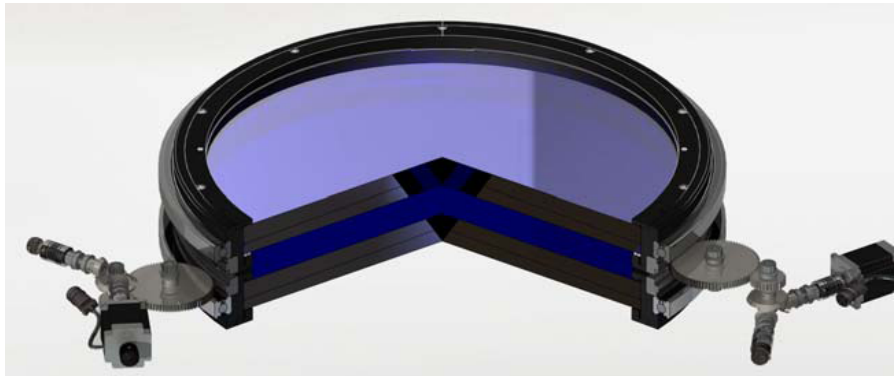

***Figure 4.3:*** *The WIYN One Degree Imager ADC is similar to that envisioned for DESpec. This ADC has a diameter of 635 mm. The prisms are rotated using a pair of encoded stepper motors.*

4.A.3 R&D

Present R&D for the optics is aimed at understanding the optimal wavelength range for the science, optimizing the focal surface, and working out details and finalizing choices of materials for the ADC and the new elements at the focal plane. The ADC itself will be made from four elements that need to be glued together (or otherwise held) in pairs; this has to be done carefully, avoiding bubbles or other defects. A mechanism that performs the anti-co-rotation of the elements needs to be specified and designed.

**4.B The DESpec Fiber Positioner**

The DESpec Wide Field Corrector optics will provide a telecentric, spherical, convex focal surface with radius of curvature ~8m and diameter 450 mm (almost the same as DECam). The fiber positioner must support ~4000 actuators across this surface, leading to a required separation between actuators (the pitch) of ~6.75 mm or less. The positioner must move the tips of the optical fibers to predetermined positions, with an accuracy of a few microns, at any RA/dec and possibly while slewing the telescope, within tens of seconds; it must then hold the fibers in place for the length of the exposure while the telescope tracks the field. It must minimize throughput and etendue losses, and it must allow the greatest possible flexibility and transparency in the allocation of fibers to targets. It must also gather the fibers into bundles that run to the spectrographs, and provide the required strain relief in doing so.

There are two general classes of fiber positioners that were considered for DESpec: the "Twirling Post" and "Tilting Spine" designs. The Twirling Post (or Cobra) design was developed for LAMOST (Wang 2000), is under development at Caltech/JPL, and is proposed for both the Subaru Prime Focus Spectrograph (Sumire, Fisher, et al. 2012) and BigBOSS (Schlegel, et al. 2011). It has the advantage that telecentricity and focus could be better maintained for all fiber positions. However, there is relatively little overlap between the actuator patrol areas compared to the "Tilting Spines," and the complexity of the design means that pitches as small as required for DESpec have not yet been achieved, and would require a substantial R&D effort.

The Tilting Spine design was developed by the AAO for the Echidna fiber positioner for FMOS on the Subaru Telescope (Akiyama, et al. 2008, Figure 4.4). It is based on a single piezoceramic actuator generating a stick-slip motion for a magnetically attached ball bearing. It allows for very small pitch between actuators (e.g., 7.2 mm for FMOS), very small separations between targets (~0.5 mm), and at least 3-fold overlaps between the patrol areas of different actuators. It has the intrinsic disadvantage of introducing telecentricity and focus errors into the beam entering the fibers. The AAO has developed and simplified the design greatly since FMOS, resolving an issue with magnetic crosstalk, while improving robustness and uniformity.

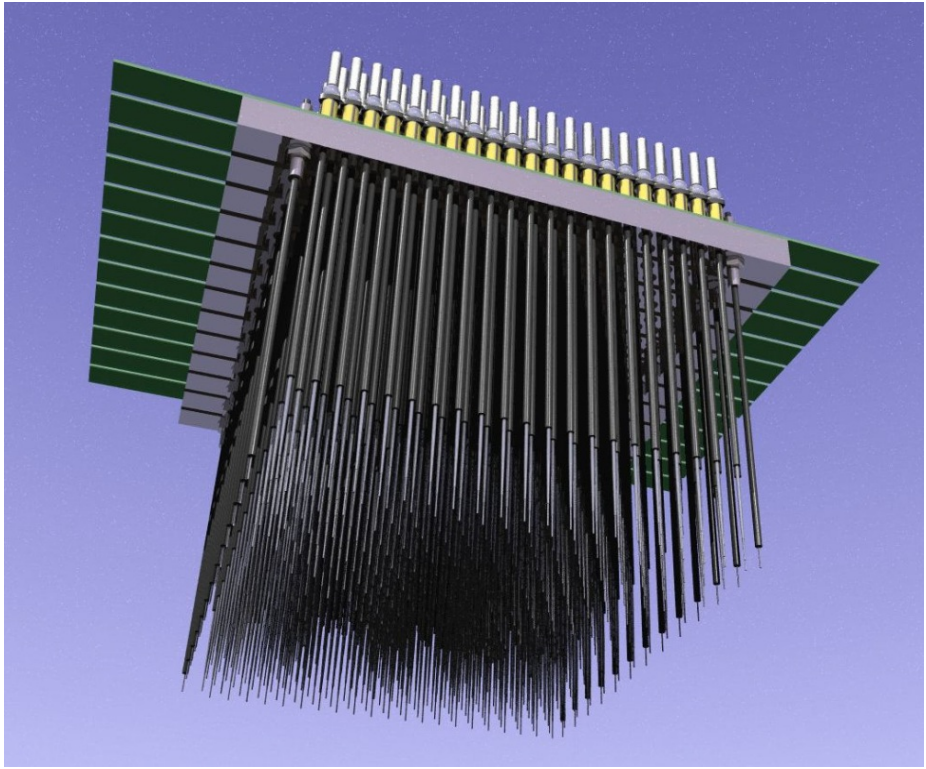

*Figure 4.4:* *400-fiber Echidna positioner for FMOS on the Subaru telescope.*

For DESpec, the AAO has designed and prototyped the 'MOHAWK' positioner concept (Saunders, et al 2012, Figures 4.5 - 4.7). This has 4000 actuators on a 6.75 mm pitch over the 450 mm diameter curved focal surface, arranged in modules for ease of construction and maintenance. The spine length has also been lengthened to 250 mm (tip to pivot), leading to maximum throughput losses from defocus and non-telecentricity of just 6.3%. No significant new technical risks have been identified over physically existing designs. The AAO is continuing to develop the design, to push down to even smaller pitches of 6 mm or less: this would allow 4000 fibers to be accommodated within a hexagonal array entirely within the 450 mm focal diameter, allowing the sky to be tiled much more efficiently.

The positioner requires a metrology system, to measure the back-illuminated fiber positions during reconfiguration to the required accuracy (5-10 microns). We propose a single large-format long-focal length camera, positioned on-axis within the Cassegrain tube. Reconfiguration is simultaneous for all fibers, but up to 7 positioning iterations are required, giving a total configuration time of 15-30 seconds. Back-illumination of the fiber-slit is required within each spectrograph, which may preclude read-out while repositioning.

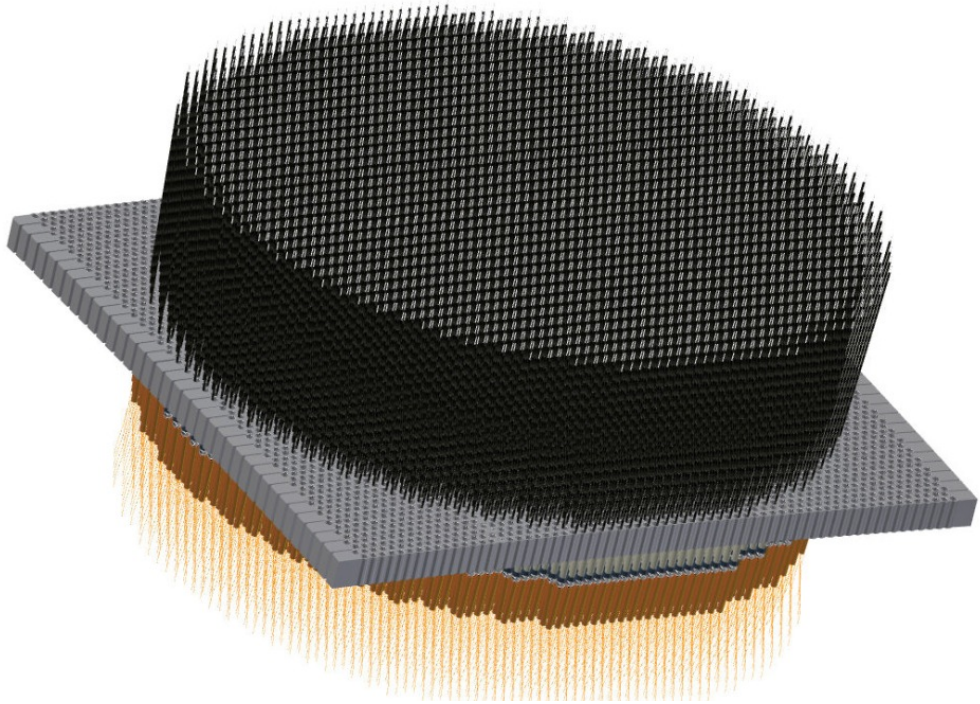

*Figure 4.5:* *Complete Mohawk actuator array, with 4000 actuators, arranged into 41 modules, filling the 450mm diameter curved focal surface (like the staves of a barrel). All science actuators are identical and all module bases are identical. Guiding, acquisition and fiducial actuators would be positioned in some of the available spaces outside the main array.*

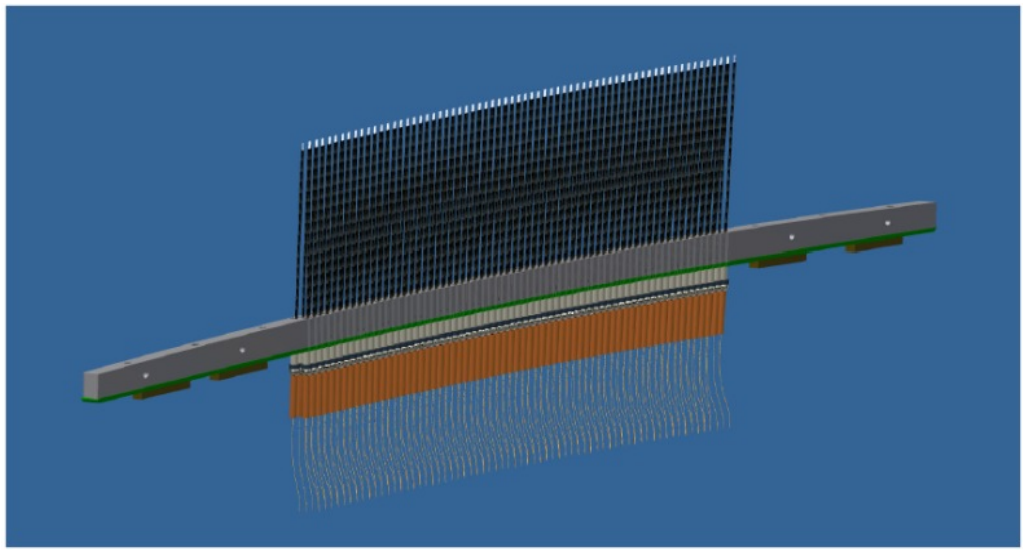

*Figure 4.6:* *A single curved MOHAWK module fully populated with actuators, also showing the multilayer circuit board bringing power to the actuators (green) and electrical sockets for connecting this to the main circuit boards (brown).*

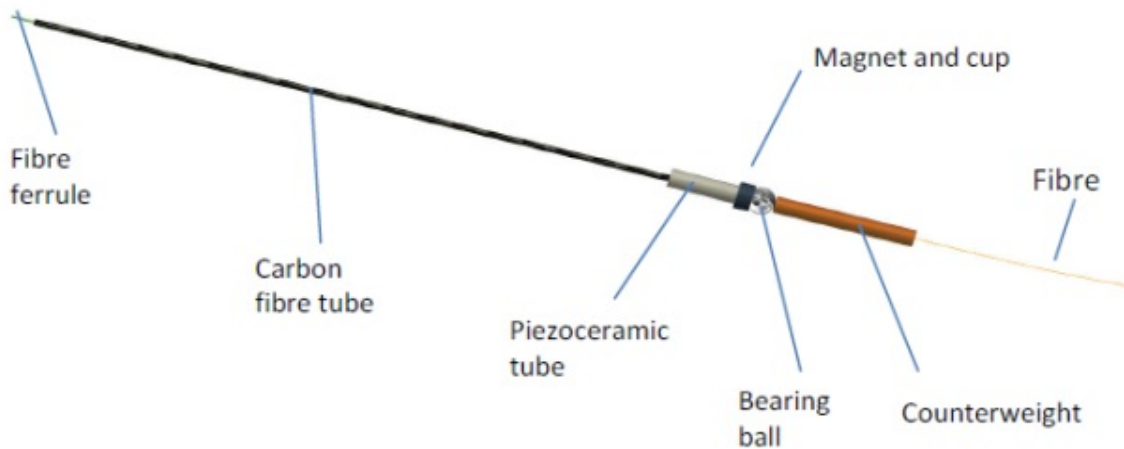

***Figure 4.7:*** *A single MOHAWK actuator. The piezoceramic tube is cemented and soldered to the module base, and has four electrodes in quadrants along its length. When supplied with a sawtooth waveform, the ball moves with sub-micron steps, allowing the spine tip to be positioned anywhere within the patrol radius. The actuator has just 7 parts in total, and only simple machining and assembly of off-the shelf items is required.*

Guiding, acquisition and focus capability are all accomplished by having separate guiding and acquisition actuators, arranged mostly outside the science array. These actuators would be identical to the science actuators, except that they would contain coherent fiber bundles of diameter ~5" and ~20" respectively. The outputs would be monitored by a single low-noise integrating video camera.

### 4.C DESpec Unit Spectrographs and Optical Fibers

DESpec spectrographs (Marshall 2012) will separate the light into component wavelengths and focus it onto CCDs. The key elements driving the optical design of the spectrographs are the wavelength range, the required spectral resolution, and the diameter of the optical fibers carrying the light from the focal surface. To accommodate 4000 fibers, each of 10 spectrographs must accept ~400 fibers. There are several options for the physical location of the spectrographs: they could be mounted off-telescope or, if sufficiently small and light-weight, they could be mounted near the top of the telescope and arrayed around its upper ring to minimize fiber length and thereby increase the throughput of the instrument. The off-telescope option for the location of the spectrographs has the feature that the fiber positioner provides a new versatile facility for CTIO. The part of DESpec up to the DESpec spectrographs can be used as-is (perhaps with the upgrade of adding an ADC) to feed other spectrographs that the observatory or the user community may wish to install, enabling a generic wide-field spectroscopic capability in the southern hemisphere. Off-telescope spectrographs can be maintained and upgraded without disrupting telescope operations.

As we are still exploring the science and survey requirements at this time, we are considering two spectrograph designs. The first design is a single-arm spectrograph with a wavelength range 550<λ<950 nm. The second is a two-arm (dichroic) spectrograph in which the blue side has wavelength range 480<λ<780 nm and the red side covers 750<λ<1050 nm.

4.C.1 Single-Arm (VIRUS) Spectrographs

The DESpec unit single-arm spectrograph optical design is based on VIRUS, the instrument being built for the HETDEX survey (Hill, et al. 2008) on the 11-meter Hobby-Eberly Telescope in Texas. HETDEX requires 200 of these spectrographs, so they must be inexpensive and simple enough to produce in bulk. The VIRUS design (Hill, et al. 2010, Marshall, et al. 2010) is a high-throughput, single-arm spectrograph. Figure 4.8 shows the optical design for VIRUS, which can be easily optimized for the DESpec wavelength regime and spectral resolution.

The VIRUS instruments are composed of the main body, housing the collimator and the disperser of the spectrograph, and a vacuum vessel that houses a Schmidt-type camera. A volume phase holographic (VPH) grating acts as the dispersing element. The collimator unit consists of a spherical collimator mirror, folding flat mirror, and the VPH grating. The Schmidt camera consists of an aspheric corrector lens that serves as the Dewar window, a spherical camera primary mirror, an aspheric field flattener lens, and a CCD detector that is held in the center of the camera on a spider assembly. DESpec, like VIRUS, is a set of simple, fixed spectrographs with no moving parts, making it inexpensive, quick to assemble, and easy to mount on or near the telescope.

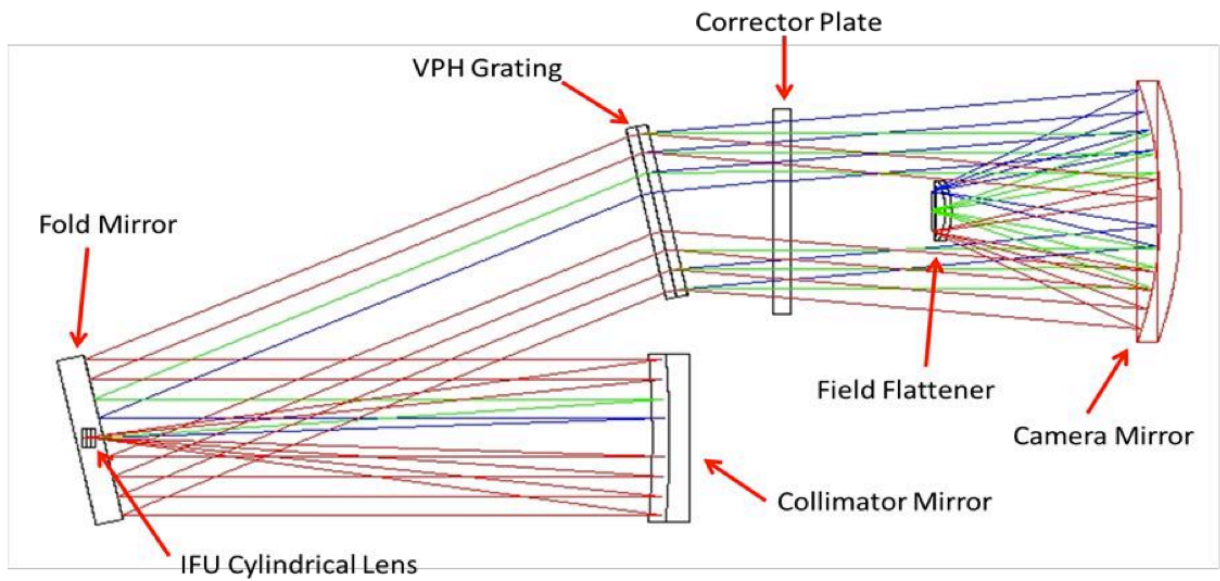

*Figure 4.8:* *Optical layout of one of the VIRUS unit spectrographs. Light enters on the left and is reflected by the collimator mirror and fold mirror before it passes through the VPH grating. The CCD is just to the left of the field-flattener lens. The overall length of the VIRUS is 0.75m.*

With the single-arm spectrograph design we use 100 μm diameter (1.75") fibers with a 2.6 pixel/fiber resolution. That requires an f/1.3 camera (HETDEX VIRUS has f/1.33). We use 20 DECam red-sensitive, fully-depleted CCDs, two for each spectrograph. Each spectrograph has 400 fibers, 200 per CCD, with spectrum along the 4k direction. Table 4.1 lists some of the parameters for the single-arm spectrograph.

| Parameter | Single-Arm |
|---|---|
| **Fiber Diameter** | 100 μm (1.75") |
| **Wavelength Range (nm)** | $550<\lambda<950$ |
| **CCD** | DECam 2kx4k |
| **Resolution ($\Delta\lambda$ nm/res. el.)** | 0.263 nm |
| **# pixels/fiber** | 2.6 |
| **Camera f/#** | f/1.3 |
| **Camera Type** | Reflective (VIRUS) |

*Table 4.1.* *The one-arm spectrograph concept.*

4.C.2 Two-Arm Spectrographs

The two-arm spectrograph design enables an increase in the wavelength range and improves the spectral resolution because we spread the light out over twice as many pixels (CCDs).

| Parameter | Blue Side | Red Side |
|---|---|---|
| **Fiber Diameter** | 100 μm (1.75") | |
| **Wavelength Range (nm)** | $480<\lambda<780$ | $750<\lambda<1050$ |
| **CCD** | E2V | DECam 2kx4k |
| **Resolution ($\Delta\lambda$ nm/res. el.)** | 0.228 nm | 0.228 nm |
| **# pixels/fiber** | 3 | 3 |
| **Camera f/#** | f/1.5 | f/1.5 |
| **Camera Type** | refractive | |

*Table 4.2.* *An example of a two-arm spectrograph with the same resolution (in nm) on the red side as on the blue side. The break at about 760 nm is to separate the two spectra at the location of a strong sky absorption feature.*

Parameters for a two-arm spectrograph covering the wavelength range $480 < \lambda < 1050$ nm are shown in Table 4.2. The spectral resolution on both sides is 0.228 nm, sufficient to easily separate the redshifted 3237A O II doublet. An example of a spectrograph of approximately this design was proposed for WFMOS. It is a high-throughput, 2-arm spectrograph with all-refractive optics and VPH gratings (Smee, et al. 2006). Another example of this type of design is the conceptual design proposed for the GMACS wide-field, multi-object optical spectrograph for the Giant Magellan Telescope. GMACS will use a fully refractive spectrograph (Marshall, et al. 2011) divided into a blue and red channel by a dichroic. VPH gratings disperse the light. Five (six) lenses focus the red-side (blue-side) of the light onto CCDs, which are at the end of each optical train. The DESpec version would be scaled down to an appropriate physical size. The benefit of a design such as GMACS is the maximal throughput of the system, in part because the beam is not occulted by a CCD in the center of the camera, higher efficiency from the two VPH gratings, and the increased wavelength.

We have adopted the two-armed spectrographs as our reference design and retain the one-armed spectrographs as a possible de-scope option. The larger spectral range (480 – 1050) afforded by the two-armed option allows, for example, both O II 372.7 and H alpha 656.3 to appear within the spectral range for all redshifts between 0.3 and 0.6, which will enhance the certainty of redshift measurement for weak-lined faint objects. The factor-of-two increase in pixels for the two-armed option translates into roughly the same increase in science reach, yet the cost of the two-armed option is much less than twice that of the one-armed option.

4.C.3 Optical Fibers

The light is carried from the fiber positioner to the spectrographs in optical fibers. The diameter of the fibers depends on the expected source flux distribution and the sky background and should be chosen to maximize the S/N ratio of a sky-dominated object spectrum.

The median delivered point-spread-function (PSF) of the Mosaic-II prime focus camera on the Blanco has been 0.9", and with upgrades related to the installation of DECam it may be a bit better. The plate scale with the DECam corrector is 0.27" per 15-micron pixel. We have initial results from a study that optimized the fiber diameter for a flux-limited survey in the presence of a dominant sky background. The calculations start with a magnitude limit and compute the total number of galaxies per sq. degree at this limit. Next a fiber diameter is chosen, and the rate at which spectra are collected to a pre-determined S/N ratio as a function of galaxy magnitude and radius is calculated. The distribution of galaxy radii and magnitudes comes from the COSMOS simulation (Jouvel, et al. 2009), as described in Sec. 3. Because large galaxies tend to have low surface brightness, they are the most difficult spectroscopic targets. The rates are calculated using a weighted average over a distribution of CTIO-measured seeing. A contribution of 0.6" from the optics PSF is included. For magnitude limits in the range 22 to 24, the optimal fiber size is in the range 1.8 to 2.0 arcsec. However, the rate at which redshifts are collected depends only weakly on fiber size: the range in fiber sizes that have success rates within 10% of the peak rate correspond to diameters ranging from 1.5 to 2.4 arcsec. The intersection of these ranges that is near-optimal for all limiting magnitudes is 1.7 to 2.1 arcsec. Because the optimization has determined that the fiber diameter is a soft minimum, we expect that the final diameter can be selected based largely on other considerations, such as spectrograph resolution or fiber positioning accuracy.

The input f-ratio, f/3, is ideal for high-efficiency in capturing the light and for high throughput and low focal ratio degradation (Murphy, et al. 2008). The length of the fibers is 10 to 30 meters and depends on where we mount the spectrographs. Figure 4.9 shows fiber throughput versus fiber length for six different wavelengths from 500 to 1100 nm in a typical fiber.

4.C.4 R&D

R&D in this area will lead to a design for the DESpec unit spectrograph and its optics. The choice among designs will come from the science and survey requirements. The R&D will optimize the fiber diameter (for instance, sky-background in some wavelengths suggests smaller fibers improve the signal-to-noise), fiber length, light throughput, and signal-to-noise ratio of the observations. We would also determine how many fibers could be read out on each CCD. An increase here could decrease the number of spectrographs by 30%, and therefore, the total cost. In addition, there is substantial R&D being performed around the world on concepts (e.g. Content & Shanks 2008; Content, et al. 2010) for multi-object spectroscopic instruments on wide-field telescopes. We are watching these efforts carefully in case a sensible technical alternative for the DESpec arises.

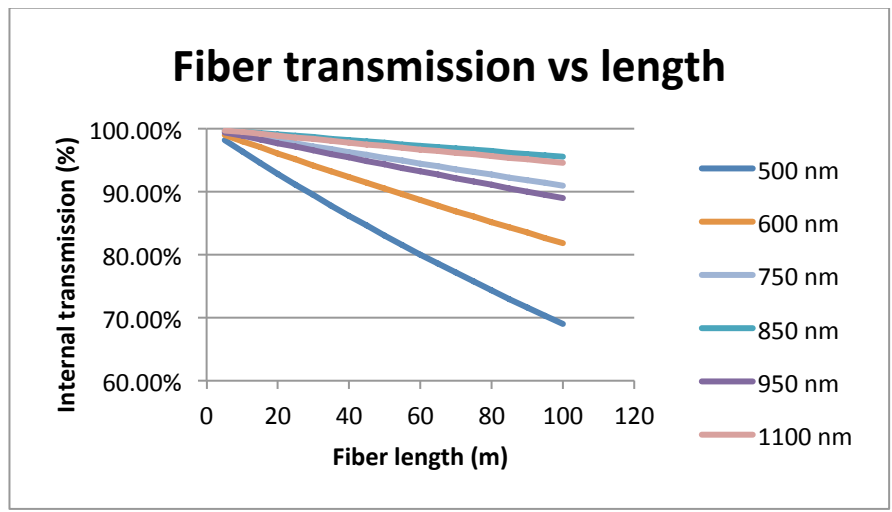

*Figure 4.9:* *Fiber throughput versus fiber length at six different wavelengths for Polymicro broadband fibers. This assumes optimal f-number, which is f/3 to f/4. Note that for shorter wavelengths the fibers have a lower throughput, which gets progressively worse for λ<500 nm.*

## 4.D CCDs and Readout

4.D.1 CCDs

The light will be dispersed by the spectrographs onto CCDs. DESpec will use 2k x 4k backside-illuminated, red-sensitive CCDs designed by LBNL, for either the one-arm spectrograph or for the red side of the two-arm spectrograph. These CCDs have high quantum efficiency (QE) at near infrared wavelengths. They are 250 microns thick and attain good (~5 micron) dispersion characteristics from a 40V substrate bias. The 4-side buttable CCD package is suitable, so existing spare, tested, packaged, science-grade DECam CCDs (Estrada, et al. 2010) can be used on DESpec, providing a significant cost saving, although we include their costs for the purpose of making cost estimates at this time. Fig. 4.10 shows the quantum efficiency of 3 DECam CCDs. The blue side of the two-arm spectrographs could also use DECam 2kx4k CCDs. E2V 2kx4k devices would also work and have QE that is a little better around 500 nm.

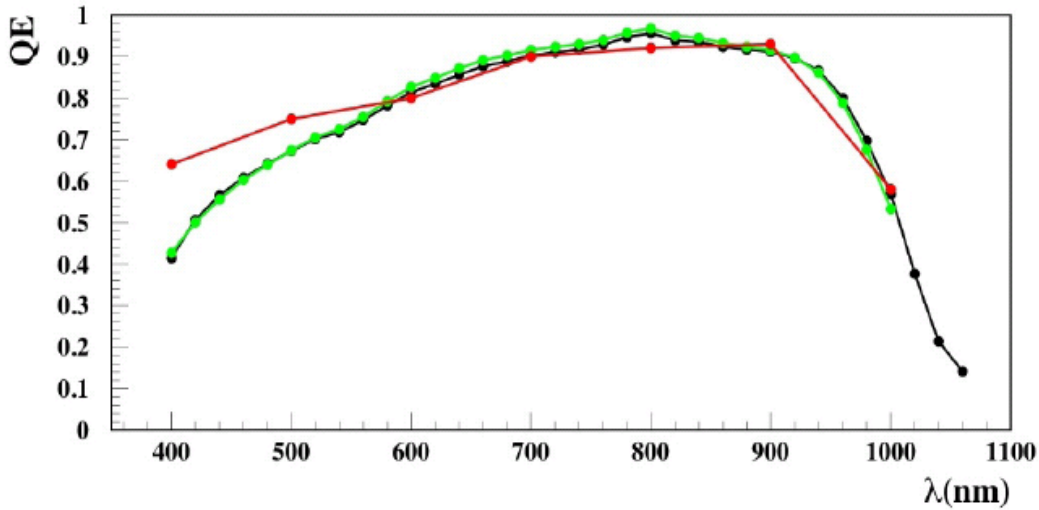

*Figure 4.10:* *The absolute QE of three typical CCDs produced for the Dark Energy Camera.*

The DECam CCD readout electronics can read out the 2kx4k CCD in ~17 (45) seconds with ~7 (3) electrons/pixel noise. They could be modified relatively easily for the E2V CCD if necessary. Even 3e- read noise is somewhat larger than desirable for faint-object spectroscopy as these resolutions (e.g., Saunders, et al. 2012), but ongoing R&D at Fermilab on low-read-noise techniques (e.g., Estrada, et al. 2012) is already achieving acceptable readout noise of 1-2e- at the required speeds.

## 4.E Interchangeability with DECam

The Dark Energy Camera has been designed for efficient installation and removal from the Prime Focus Cage. The time required to change between DECam and DESpec (total of one or two days) may be limited by the warm-up time for the DECam imager. Instruments at the f/8 focus can be used for time when the prime focus is not available. Figure 4.11 shows the camera installation fixture positioned in front of the Prime Focus Cage, mounted on the Telescope Simulator at Fermilab (Diehl, et al. 2010). To change from DECam to DESpec, one tilts the Blanco over to the northwest platform and uses the camera installation fixture to remove DECam. DECam is then stowed off of the telescope with its Dewar window, the camera's final optical element (C5), in place. DESpec, which will have been stowed either off-telescope or on the telescope structure, is connected to a similar installation fixture for inserting into the cage. DESpec would be approximately the same weight as DECam; differences in weight would be corrected for by adjustment of the prime focus cage counterweights.

The DECam Project has supplied CTIO with a platform that allows easy installation and removal of the filter-changer and shutter. The platform is with the telescope at the North position. DESpec's ADC and (the same) shutter will be installed and removed from the same platform using a similar procedure.

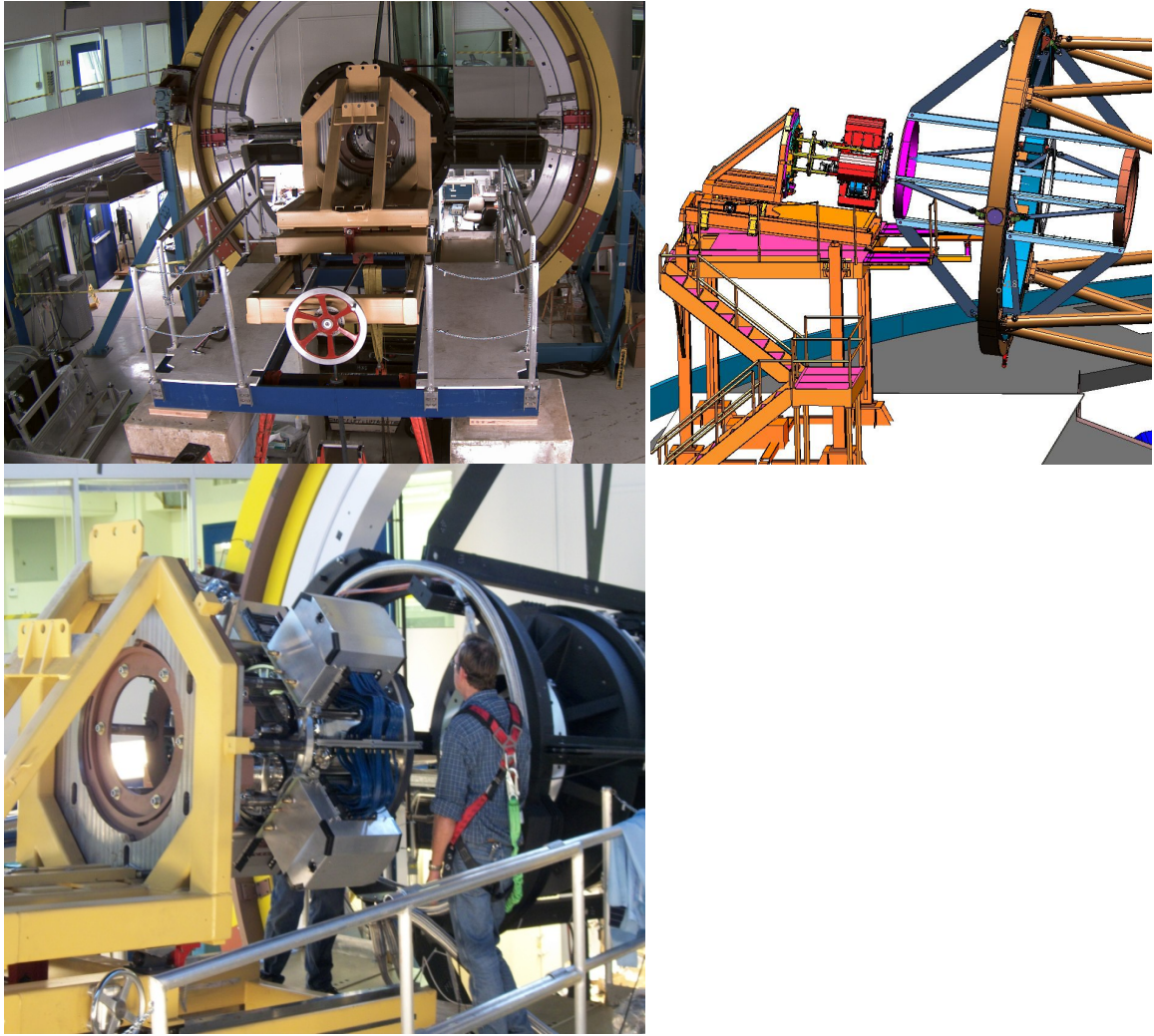

***Figure 4.11:*** *The camera installation fixture (in the foreground at top left, schematic at top right, in close-up at bottom) at Fermilab. Bottom image shows the installation fixture being used to mount the camera in the Prime Focus Cage (black, at right). The cage itself is attached by fins to the white and yellow rings of the telescope simulator; the inner white ring has the same dimension as the ring at the top end of the Blanco. The simulator was used to test DECam in all configurations it will encounter on the telescope as well as the mounting and dismounting procedures.*

## 4.F The Blanco Telescope

The Blanco 4m telescope was named after Dr. Victor M. Blanco (1918-2011), an astronomer from Puerto Rico, who was the second Director of CTIO (1967-1981). The Blanco is an equatorial mount telescope with a 150" diameter primary mirror. The primary mirror and prime focus cage are mounted on a Serrurier Truss (Abdel-Gawad, et al. 1969). The telescope was commissioned in 1974 (Blanco 1993). Until 1998, it was the largest telescope in the Southern Hemisphere. It has a high-quality Cervit primary mirror that provides excellent image quality. An extensive set of improvements made a decade ago included replacement of the passive primary mirror supports with an active system and alterations to the telescope environment to improve the air flow and remove heat sources. More recent improvements made for the Dark Energy Camera include replacement of the primary mirror's radial supports, which has substantially reduced motion of the primary in its cell, and an upgrade of the telescope control system. The cover figure shows a photograph of the Blanco 4m telescope with the new DECam prime focus cage installed. The same cage, optics, and hexapods would be used for DESpec.

## 4.G Technical Summary

We have described technical solutions to the DESpec design and demonstrated that there are no significant technical risks. For the major new systems, we have provided examples that have already been designed and built for existing or near-term instruments. While these are not final technical choices, they do represent solutions that satisfy our project scope and goals with minimal R&D. The technical, schedule, and cost risks of the project appear to be manageable.

We have estimated that this instrument is approximately the same scale project as DECam and can be built at roughly the same cost and on similar timescale. Members of this team have completed the $35M DECam Project on budget and on schedule. Our experience with DECam has guided our estimate for the DESpec cost and schedule. The reference design of DESpec has 4000 fibers, 10 double spectrographs, and no ADC. An optical design for the corrector exists that allows an ADC to be added later, which would be a good upgrade option in the future (e.g.. to observe LSST fields at higher airmass). There are a number of plausible descope paths, for example a smaller number of fibers (and spectrographs) and single-arm spectrographs (for which we would require no additional CCD's). For the double-arm spectrograph reference design, our DESpec cost estimate includes materials and construction at $27M, plus 47% contingency ($13M) (relatively high, not because of perceived risk, but because the design and scope are still under development and will be optimized based on the scientific case). Including the contingency, the total DESpec construction cost is expected to be no more than $40M. Adding an ADC would increase this number by about $1M, and adopting single-arm spectrographs would decrease the total by about $9M, including contingency. We expect that substantial costs will be defrayed through the contribution of university and international collaborators, as was the case with DECam.

## 5. Project Summary

DESpec offers an extraordinary opportunity for advances in cosmology and fundamental physics, and the instrument enables community access to wide-field, deep, multi-object spectroscopy in the southern hemisphere in the LSST era. Much of the required hardware is already in place with DECam. The minimal scope of the DESpec project is specified by the technical requirements to constrain cosmological parameters, but higher performance can be achieved with straightforward enhancements to the reference design. As this white paper was being finalized for distribution, the report of the NSF Astronomy Division Portfolio Review Committee appeared, which endorses the importance of wide-field, highly multiplexed spectroscopy for cosmology and galaxy studies.

**Appendix: DESpec and BigBOSS**

As noted above, the combination of DESpec in the south and BigBOSS in the north would enable spectroscopic surveys over the entire sky, maximizing the DE constraints that could be pursued from the ground. Here, for context we briefly compare and contrast the DESpec and BigBOSS designs. BigBOSS is a proposed 5000-fiber spectrograph system for the Mayall 4-meter telescope at Kitt Peak National Observatory in Arizona (Schlegel, et al. 2011). The Mayall and Blanco telescopes are essentially identical mechanically, and both observatories are operated by NOAO. The BigBOSS design calls for a larger field of view (FOV) than DESpec (7 vs. 3.8 sq. deg.), requiring several larger optical corrector elements and entailing an entirely new prime focus cage, active alignment system, and set of corrector lenses for the Mayall. The BigBOSS robotic fiber system would be similar to that used for the LAMOST project in China; it has a larger pitch (inter-fiber physical separation) that is technically less challenging to achieve than DESpec's. BigBOSS would employ multi-arm spectrographs to span a broader range of wavelengths (extending down to 340 nm, to access the Lyman-alpha forest along QSO sight-lines) at slightly higher spectral resolution than currently envisioned for DESpec. BigBOSS is optimized to probe Baryon Acoustic Oscillations (BAO) to redshifts $z>1$ and would select spectroscopic targets mainly from WISE (infrared), Palomar Transient Factory (PTF), and PanSTARRS imaging, which are shallower than DES. Given its location at relatively high latitude, BigBOSS would be able to target only ~600 sq. deg. (~1/8) of the nominal DES survey area and up to ~several thousand sq. deg. of the LSST footprint, so it would not take full advantage of the synergy between weak lensing (DES, LSST) and redshift-space distortions from the same volume (Secs. 2, 3) nor would it be able to follow up the majority of the DES and LSST target area, nor overlap with South Pole Telescope survey fields. The deep, multi-band, precisely calibrated photometry from DES and LSST should in principle enable more efficient spectroscopic targeting, a topic requiring further study. On the other hand, the DES and BigBOSS collaborations have jointly explored the possibility of increasing the area overlap between DES/DECam imaging and BigBOSS spectroscopy through increased DES survey time in the celestial equator regions; an overlap of up to several thousand sq. deg. is possible but would require DECam imaging time beyond the 525 nights allotted to the DES project.

In order to reach the density of galaxy targets optimal for BAO studies, BigBOSS would revisit each field several times. DESpec, with a higher density of fibers on the sky, would reach the same target density in fewer visits. However, since DESpec has a smaller FOV, the two instruments would in fact have comparable speed (number of targets per unit area per unit time) for a BAO-optimized survey. BigBOSS proposes to measure ~20 million galaxy redshifts over 14,000 sq. deg. in 500 nights. More conservatively assuming longer cumulative exposure times to ensure redshift success, we estimate that DESpec could measure ~8 million redshifts over 5000 sq deg. in ~350 nights and by extension ~23 million redshifts over 15,000 sq. deg. with DES+LSST imaging in ~1000 nights. DESpec would enable many other follow-up programs over the entire 20,000 sq. deg. LSST survey area as well. Given the site conditions, DESpec would be expected to have a higher fraction of useable nights than BigBOSS (~80% vs. ~65%) and better seeing (Els, et al. 2009).

The BigBOSS proposal of Oct. 2010 responded to an NOAO Announcement of Opportunity for a new instrument for the Mayall telescope; in Jan. 2011, the Large Science Program panel that reviewed the proposal recommended that NOAO work with the BigBOSS team to further develop the proposal, retire key risks to the project, and bring it to a state of readiness for submission to DOE. As indicated above in Sec. 1, NOAO currently has no plans to issue an announcement of opportunity for a new prime-focus instrument for the Blanco, but we note that DESpec already has many of its optical and mechanical parts installed on the telescope, and the construction phase is likely to be relatively fast once it begins.